\documentclass[lettersize,journal]{IEEEtran}
\usepackage{amsfonts}
\usepackage{amssymb}
\usepackage[ruled,vlined]{algorithm2e}
\usepackage{mathtools}  
\usepackage{amsthm}
\usepackage[]{algorithm2e}
\usepackage{subfigure}
\usepackage{tabulary}
\usepackage{footnote}
\usepackage[export]{adjustbox}
\usepackage{booktabs}
\usepackage[justification=centering]{caption}
\usepackage{comment}
\newtheorem{remark}{Remark}
\usepackage{cases}
\usepackage{cite}
\usepackage{multicol}
\usepackage{xcolor}
\usepackage{graphicx}
\graphicspath{{./}{figures/}}

\usepackage{stfloats}
\usepackage{cuted}
\usepackage{lipsum}  
\DeclareMathOperator*{\argmax}{argmax}
\DeclareMathOperator*{\argmin}{argmin}
\usepackage{subfigure}

\newtheorem{lemma}{Lemma}
\newtheorem{theorem}{Theorem}

\title{Multi-Access Point Coordination for Next-Gen Wi-Fi Networks Aided by Deep Reinforcement Learning} 

\author{\text{Lyutianyang Zhang}, \text{Hao Yin},  \text{Sumit Roy}, \IEEEmembership{Fellow,~IEEE}, \text{Liu Cao}
\thanks{This paper has been presented in part at the IEEE Vehicular Technology Conference  (VTC) 2020-Fall \cite{ZhangVTC2020}.

Lyutianyang Zhang, Hao Yin, Sumit Roy, and Liu Cao are with Department of Electrical \& Computer Engineering, University of Washington, Seattle, WA, USA (e-mail:\{lyutiz, haoyin, sroy,	liucao\}@uw.edu). (\emph{Corresponding author: Hao Yin})}}
\begin{document}

\maketitle

\begin{abstract}
Wi-Fi in the enterprise - characterized by overlapping Wi-Fi cells - constitutes the design challenge for next-generation networks. Standardization for recently started IEEE 802.11be (Wi-Fi 7) Working Groups has focused on significant medium access control layer changes that emphasize the role of the access point (AP) in radio resource management (RRM) for coordinating channel access due to the high collision probability with the distributed coordination function (DCF), especially in dense overlapping Wi-Fi networks. This paper proposes a novel multi-AP coordination system architecture aided by a centralized AP controller (APC). Meanwhile, a deep reinforcement learning channel access (DLCA) protocol is developed to replace the binary exponential backoff mechanism in DCF to enhance the network throughput by enabling the coordination of APs. First-Order Model-Agnostic Meta-Learning further enhances the network throughput. Subsequently, we also put forward a new greedy algorithm to maintain proportional fairness (PF) among multiple APs. Via the simulation, the performance of DLCA protocol in dense overlapping Wi-Fi networks is verified to have strong stability and outperform baselines such as Shared Transmission Opportunity (SH-TXOP) and Request-to-Send/Clear-to-Send (RTS/CTS) in terms of the network throughput by 10\% and 3\% as well as the network utility considering proportional fairness by 28.3\% and 13.8\%, respectively.
\end{abstract}
\begin{IEEEkeywords}
Wi-Fi 7, IEEE 802.11be, multi-AP coordination, channel access, proportional fairness, deep Q-learning
\end{IEEEkeywords}

\section{Introduction} 
\label{introduction}

The rapid adoption of smartphones, tablets, and high-end mobile client devices has translated into the rapid growth of {\em network traffic flux} (measured in bits/s/Hz per unit area/volume). As tracked by Cisco Annual Internet Report \cite{Cisco}, the number of Wi-Fi hotspots will grow four-fold, and the average mobile network connection speeds will triple from 2018 to 2023.

Multi-media streaming demands, e.g., 4K and 8K video \cite{Perez2019AP}, will stretch network capacity even beyond current generation (Wi-Fi 6) limits of $10$ Gbps peak capacity. To address such technology bottlenecks, the IEEE 802.11 Working Group (WG) is standardizing the next generation of Wi-Fi—referred to as IEEE 802.11be (Wi-Fi 7) or Extremely High Throughput (EHT) networks. 

\begin{figure}
    \centering
    \includegraphics[width=.50\textwidth]{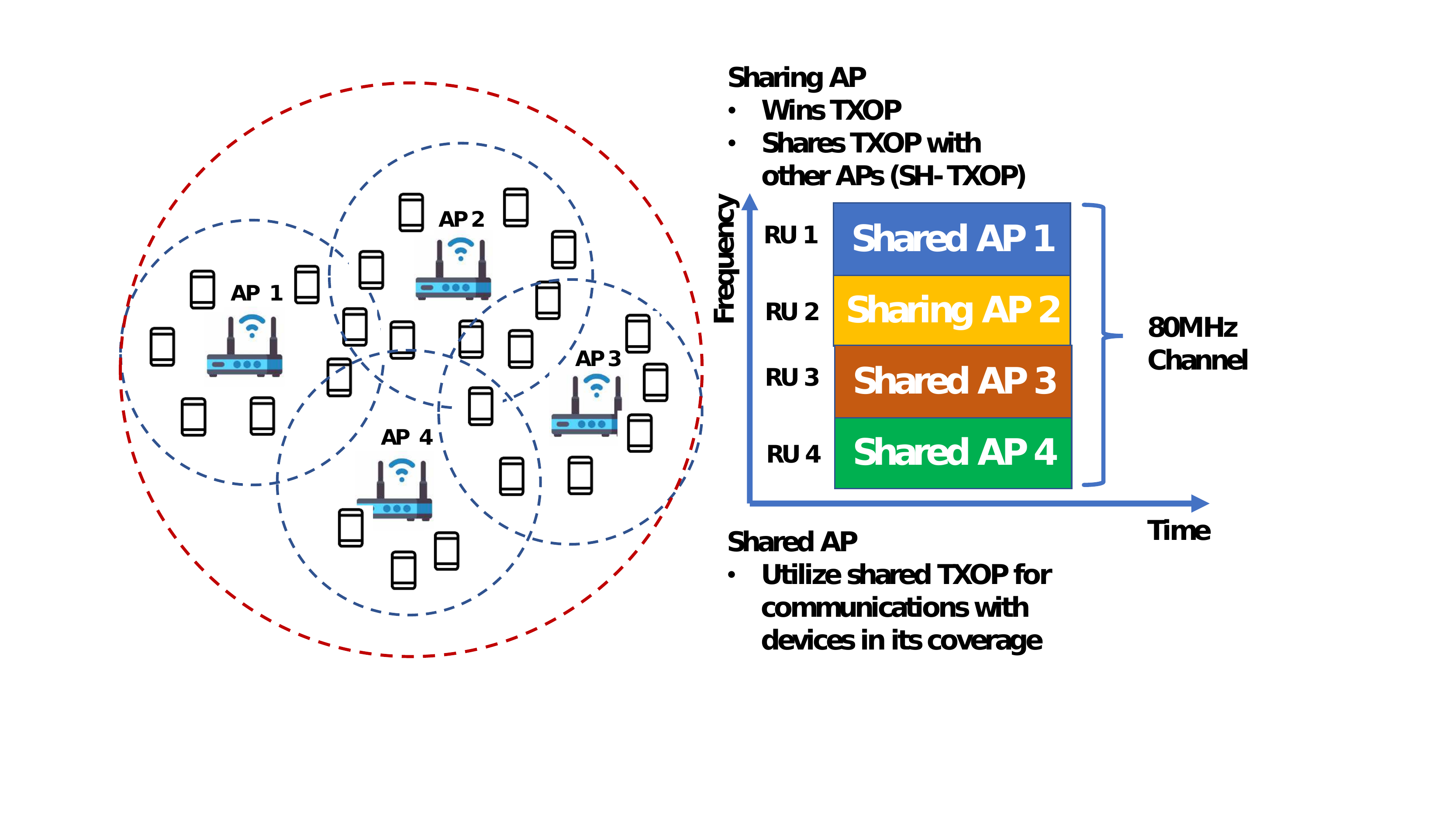}
    \caption{Multi-AP Coordination with SH-TXOP: 4 APs share a 80 MHz universal channel. Each AP has its own coverage of stations (STAs) and operates on its own primary channel. Each resource unit (RU) represents a 20 MHz channel.}
    \label{fig:wlan}
\end{figure}

Orthogonal frequency-division multiple access (OFDMA) adopted in 802.11ax has significantly enhanced the medium access control (MAC). OFDMA works on top of the legacy carrier-sense multiple access with collision avoidance (CSMA/CA) to provide extra features when access point (AP) contends for channel access, e.g., the use of trigger frame helps control the uplink transmission of stations \cite{11axTutorial} for greater efficiency in a single basic service set identifier (BSS-ID). The overlap of APs is defined as the intersection of their cells and operating frequency bands. In such a cluster of overlapping AP, channel access collision happens more frequently as the number of APs increases, especially when each AP has no prior knowledge of other APs' channel accessing policies. Thus, enabling some degree of collaboration among neighboring APs will permit more efficient utilization of the limited time and frequency resources, i.e., lower collision probability, higher network throughput. To this end, the next-generation standard 802.11be (Wi-Fi 7) introduces some additional features such as multi-AP coordination to further improve aggregate throughput in dense overlapping layout scenarios \cite{11beMultiAP}. 

The emphasis of EHT WG is on {\em aggregate throughput in dense networking scenarios} and hence - building on the numerous physical layer (PHY) advances made in 802.11ac/ax - notably new control frames for coordination among APs that requires information exchange among the APs belonging to the coordinated AP set. Besides, the new 6 GHz bands opened up in the US and Europe are new green fields for the future Wi-Fi 7 standard, which gives more freedom in architecture and protocol design. In Fig.\ref{fig:wlan}, an example of the current 11be structure with Shared Transmission Opportunity (SH-TXOP) operation is introduced \cite{shtxop}. Four APs simultaneously operate on the shared 80-MHz bandwidth. Distributed coordination function (DCF)  applied in SH-TXOP allows all APs to contend for channel access. In this example, AP 2 successfully gains the TXOP and becomes a sharing AP. AP 2 then collects information about the channel status and the traffic backlog from shared APs in the candidate shared AP set. Afterwards, the sharing AP will share the wide-band TXOP with the shared APs. Each AP must have a primary channel to operate in the dense overlapping network so that it can contend for TXOPs for the communication within its own coverage. Note that primary channels allocated to different APs are not necessarily the same. In such a scheme considering collaboration, there is no frequency overlap among four channels because each AP operates within its own allocated channel independently. The bandwidth of each channel occupied by a shared AP or sharing AP is $20$ MHz. 

\begin{figure}
    \centering
    \includegraphics[height=0.3\textwidth, width=.4\textwidth]{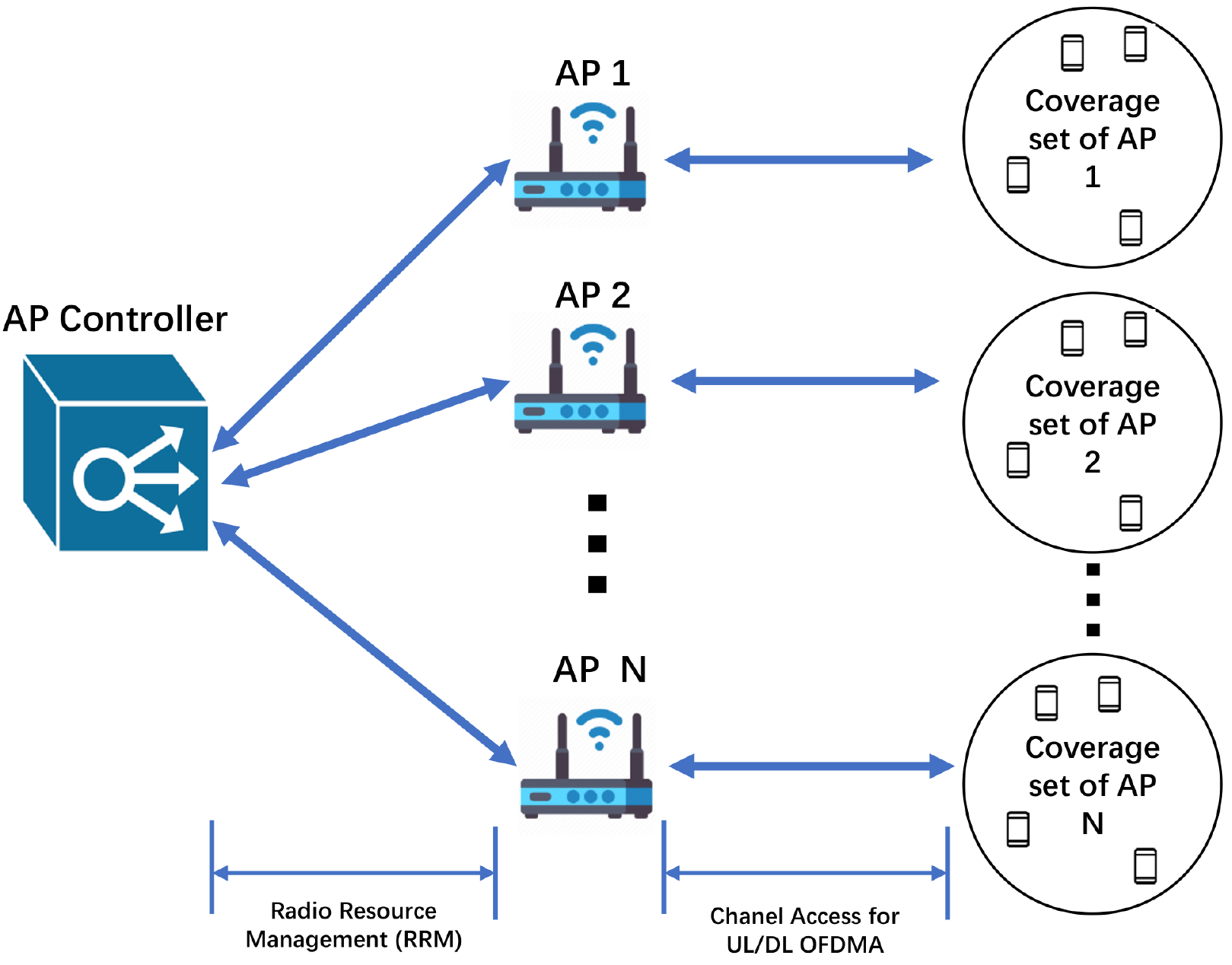}
    \caption{Proposed Architecture for Multi-AP Coordination.}
    \label{fig:APC}
\end{figure}

\subsection{Multi-AP Coordination Architecture and Related Work} 

DCF with CSMA/CA is a traditional MAC protocol for channel access in Wi-Fi networks. It has a long history of analysis using Markov models \cite{kwak2005performance,bianchi2000performance} developed originally for a single (isolated) cell (e.g., single home networks) with saturated nodes.

In \cite{ahn2020novel}, a novel multi-AP coordination transmission scheme is proposed for 802.11be. TXOP in the proposed scheme is acquired after the AP completes its backoff procedure. Then, it performs a wide-band
transmission if the secondary channel is idle during the point coordination function (PCF) inter-frame spacing (PIFS). This contention-based method for TXOP is similar to DCF Request-to-Send/Clear-to-Send (RTS/CTS) \cite{bianchi2000performance}. Then, the sharing AP sends an Announcement Trigger Frame (ATF) to the shared APs to allocate all 20 MHz channels, including the duration of TXOP and the scheduled channel information. This proposed scheme also aligns with Coordinated OFDMA (C-OFDMA). That is, each shared AP utilizes partial TXOP assigned by the sharing AP for its uplink/downlink (UL/DL) OFDMA transmission with its associated STAs.

As the backoff procedure for granting the wide-band TXOP is performed solely by the initiating AP, all responding APs should terminate the entire transmission sequence before the TXOP duration, indicated by the received ATF. However, this leads to profound performance loss as the number of APs grows larger. In \cite{ahn2020novel}, only 2 APs operate on a 40 MHz channel. The backoff procedure not followed by ACK can lead to the following situation: more than one AP may think it has won the TXOP and start sending ATF, which will lead to a collision. Since there is no feedback such as ACK, in the end, the whole TXOP is wasted because more than one AP send different RU assignment in ATF and any AP will obtain confusing allocation scheme. As a result, the collision probability increases significantly with the increasing number of APs. Moreover, it is challenging to design a feedback mechanism such as ACK for the TXOP contention method. Responding to the sharing AP by all shared APs is a huge burden to the system performance because any failure reception of the ATF to any AP will lead to re-transmission. 

A novel system architecture is thereby proposed for multi-AP coordination in 802.11be to decrease the collision probability of channel access, as is shown in Fig.\ref{fig:APC}. The centralized AP controller (APC) implements channel configuration, i.e., assigning primary channels to all APs with consideration of proportional fairness (PF). Under this system architecture, APs do not need to contend for the wide-band TXOP, and ACK is much easier to design. The core function provided by APC is called radio resource management (RRM) \cite{7727996}, which automatically monitors traffic, capacity, and reliability of operating APs. RRM can periodically reconfigure the 802.11 networks for best efficiency by performing functions such as radio resource monitoring and dynamic channel assignment. Each AP contends for TXOP on the assigned primary channel which is further utilized for the communications with its associated STAs by UL/DL OFDMA.

The suffering from the traditional DCF characterized by collision probability in dense overlapping networks also prompts the applications of state-of-the-art machine learning techniques. Deep reinforcement learning (DRL) is a machine learning technique that enables an agent to take actions in an environment aiming to maximize the cumulative rewards. Reinforcement learning (RL) principles have shown potential on optimizing resource allocation in various aspects in wireless communication, see \cite{application1,application2,application3,ding2020deep}. Nonetheless, the application of DRL in wireless networks must be made wisely, as the network utility unavoidably oscillates due to the unstable nature of RL \cite{mnih2015human}. Moreover, in the overlapping multi-AP network, the network throughput is affected by the different channel qualities, the number of users, etc. Therefore, the designed AP coordination algorithm should also account for those factors. Adversarial RL-based method \cite{kihira2020adversarial} is proposed as the solution to single-band multi-AP coordination in 11be. Deep Q-network (DQN) is investigated as an enhancement (higher utility) for CSMA in heterogeneous networks in which more than one multiple access protocol coexist \cite{yu2019deep,yu2020non}. However, this work assumes that all stations run on the same frequency band. As a more ambitious study, multiple-agent deep learning multiple access with imperfect transmission feedback \cite{yu20211} is studied. The main objective in \cite{yu20211} is to recover the lost transmission feedback between AP and STA due to imperfect channel condition. In \cite{yu20212}, multi-channel access for multiple STAs and one AP based on deep reinforcement learning is investigated. This paper concentrates on that all mobile users utilize improved channel access approaches to communicate with one AP. Our work differs from \cite{yu20211,yu20212} by considering the throughput maximization problem for multiple overlapping and collaborating APs following 11be standards in which AP is in charge of channel access for TXOP. In \cite{ZhangVTC2020}, we consider a single channel access problem for the communication between only one AP and multiple STAs. As an extended version to \cite{ZhangVTC2020}, we propose in this paper a novel system architecture that consists a centralized APC with a PF solution to multi-channel allocation problem.

By contrast, our work also differs from the above research work by investigating a new multi-band multi-AP coordination network with APC based on IEEE 802.11be, and our proposed protocol simultaneously enables TXOP contentions at different frequency bands.

\subsection{Contribution}
This paper proposes a novel coordinated multi-AP architecture and a corresponding channel access mechanism aligning with IEEE 802.11be to maximize the aggregate network throughput while preserving fairness among APs. The major contributions of this paper are listed as below:
\begin{itemize}
\item 
We propose a multi-AP system with APC as well as formulate a dynamic resource allocation and channel access optimization problem. The resource allocation process is considered as a Markov decision process (MDP). We choose the previous observation of channels and actions as the state, transmission at a channel as the action, and successful/unsuccessful transmissions at multiple channels as positive/negative rewards.
\item 
Deep reinforcement learning channel access (DLCA) protocol is proposed. For each AP in the coordinated multi-AP set, DLCA is deployed to contend for channel access. The first AP winning the contention gains the TXOP on its primary channel. The First-Order Model-Agnostic Meta-Learning (FOMAML) is then applied to DLCA to enhance the overall performance. We also develop a greedy algorithm to maintain PF among APs. 

\item Simulation results show that the performance of DLCA protocol is verified to have strong stability and outperform baselines such as SH-TXOP and RTS/CTS in terms of the network throughput as well as the network utility in dense overlapping Wi-Fi networks.

\end{itemize}

The paper is organized as follows: Sec II introduces the proposed system model aligning with IEEE 802.11be protocol. Next, in Sec III, DLCA plus greedy algorithm with FOMAML are combined and developed as the DLCA protocol for this system. Sec. IV contains a suite of performance evaluations for the proposed DLCA protocol. The comparison between DLCA protocol and baselines is implemented to demonstrate the robustness and efficiency of our proposed DLCA protocol.
\begin{table}[htbp]
    \centering
    \caption{Main Acronyms}
    \label{table:notation}
\setlength\extrarowheight{1pt}
\resizebox{.45\textwidth}{!}{%
\begin{tabular}{|p{35 pt}|p{150 pt}|}
\hline
TXOP                      & Transmission Opportunity      \\ \hline
SH-TXOP & Shared Transmission Opportunity                                                               \\ \hline
STA                       & Station                                                                 \\ \hline
RU       & Resource Unit                                                         \\ \hline
DIFS                     & Distributed Inter-Frame Space          \\ \hline
SIFS                     & Short Inter-Frame Space          \\ \hline
RTS/CTS                    & Request-to-Send/Clear-to-Send          \\ \hline

\end{tabular}}
\end{table}

\section{System Model and Problem Formulation of DLCA}
\label{sec:architecture}

In this section, we firstly introduce the multi-AP network with APC. Then the DCF RTS/CTS is introduced and the novel DLCA is proposed. We formulate each AP's dynamic resource allocation and channel access optimization problem as MDP.

\subsection{Proposed Multi-AP network}
In 802.11be system, the aggregation of $5$ and $6$ GHz spectrum allows simultaneous operation on different bands or channels (orthogonal frequency resource allocation). Our proposed network model aligning with 11be protocol is shown in Fig.\ref{fig:APC}. The coordinated multi-AP set is defined as  $\mathcal{N}=\{1,\dots,n,\dots,N\}$, and $\mathcal{F}=\{1,\dots,f,\dots,F\}$ denotes the available orthogonal channel set. Each AP can be allocated with a different channel from the available channel sets. Suppose, at the $t^{th}$ contention for TXOP, AP $n$ observes the channel states of its allocated primary channel (the $f^{th}$ channel of 20 MHz), which yields the observation vector $o^{n}_{t}(f) \in \{0,1\}$ where $0$ and $1$ denote the IDLE and BUSY channel state, respectively. Action vector is denoted as $a^{n}_{t}(f) \in \mathcal{A} \triangleq \{0,1\}$ where $a^{n}_{t}(f)=1$ represents AP $n$ contends for the $f^{th}$ channel at $t^{th}$ contention for TXOP and $a^{n}_{t}(f)=0$ represents AP $n$ does not contend for the $f^{th}$ channel at $t^{th}$ contention for TXOP. A successful transmission occurs if a sole AP occupies a TXOP. In the following section, the action and observation vector are simplified as $a^{n}_{t}$ and $o^{n}_{t}$, respectively, because the $f^{th}$ channel is implicitly linked to AP $n$ after APC has made the channel allocation decision.


\subsection{Channel Access Mechanism}
\begin{figure*}[t]
    \centering
    \includegraphics[width = 0.95\textwidth,height=0.25\textwidth]{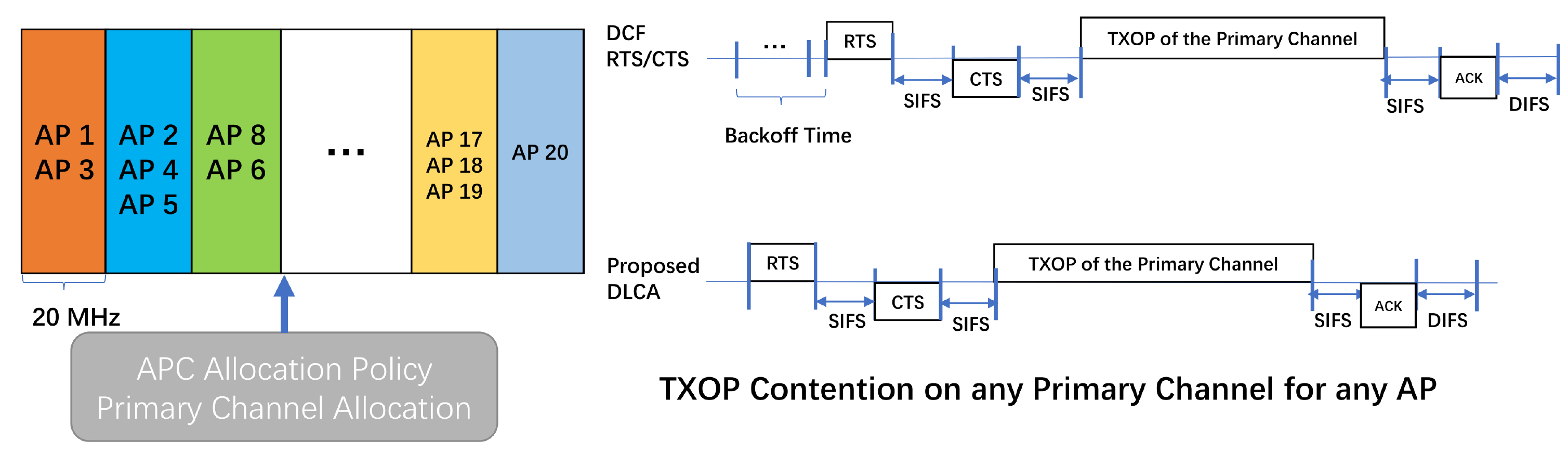}
    \caption{The primary channels of APs are assigned by the APC. $F$ channels from 5 and 6 GHz bands are available for IEEE 802.11be. Then APs in each 20 MHz band contends for TXOP.}
    \label{fig:system}
\end{figure*}

This section briefly introduces three packet mode channel access protocols that can be potentially utilized in our proposed multi-AP network with APC, including our proposed DLCA protocol. They are described as follows:
\begin{itemize}
\item

{\em Distributed coordination function (DCF) basic \cite{bianchi2000performance}}: Suppose an AP wants to occupy the TXOP on its primary channel. It waits until the channel is sensed idle for a distributed inter-frame space (DIFS). Then, a backoff process is initiated. Backoff intervals are slotted, and the discrete backoff time is uniformly distributed in the range [$0, W-1$], where $W$ is defined as the contention window size, and $CW_{min}$ represents the minimum contention window. The backoff counter is utilized for AP to decide whether to access channel at the current time slot. The backoff counter value is initialized by uniformly choosing an integer from the range [$0, W-1$]. Then, it is decremented by one at the end of each idle slot. Note that the backoff counter will be frozen when a packet transmission is detected on the channel and will be reactivated until the channel is sensed idle again for a DIFS period. The AP contends for TXOP when its backoff counter reaches zero. The ACK follows after the completion of the TXOP unless collision happens. If unsuccessful TXOP happens, contention window size $W$ is doubled after each unsuccessful transmission, up to a maximum value $CW_{max}=2^{m}CW_{min}$, where $m$ represents the largest times the contention window size can be doubled.

\item
{\em DCF Request-to-Send/Clear-to-Send (RTS/CTS)}: AP transmits a short frame of RTS to APC after the backoff counter is decremented to zero. When APC detects an RTS frame, it responds, after a short inter-frame space (SIFS), with a CTS frame. The AP can only occupy the TXOP of its primary channel if the CTS frame is correctly received. The RTS and CTS frames also carry the information of the TXOP duration to be transmitted. This information can be heard by any listening AP, which can then update a network allocation vector (NAV) containing the information that the duration of the channel being busy. Therefore, an AP can suitably delay further transmission by detecting just one frame among the RTS and CTS frames and thus avoid collision. The major difference between basic and RTS/CTS is that RTS/CTS will send a RTS and decide to contend for the TXOP only after a CTS is received from the APC. DCF basic, on the other hand, contends for TXOP without sending a RTS. Hence, when collision happens, DCF basic wastes a whole TXOP duration while DCF RTS/CTS only wastes a RTS/CTS duration. It is noteworthy that TXOP can take up to 8.16 ms and RTS/CTS only take up to 0.4 ms. Hence, DCF RTS/CTS mechanism is very effective in terms of network throughput, especially for large data load in TXOP, as it reduces the average number of wasted time slots involved in the contention process \cite{bianchi2000performance}. 


\item
{\em Deep reinforcement learning channel access (DLCA)}: In the DLCA protocol, APC periodically assigns the primary channel to each AP considering PF (as formulated in section \ref{PF}). Then, each AP senses the channel and obtains an observation from its primary channel environment, indicating the channel is BUSY or IDLE. Based on the observed results, each AP implements inference regarding the next action utilizing its trained deep Q learning model to maximize the network throughput in its coverage (as formulated in section \ref{AP}). It is noteworthy that once the AP employing DLCA decides to contend for TXOP, the similar protocol to DCF RTS/CTS in Fig.\ref{fig:system} is followed. The only difference is that DLCA has no backoff time, and it transmits RTS as long as it determines to contend for TXOP, which makes the Coordinated-OFDMA \cite{Perez2019AP} possible because each contention process has a constant duration and can be visualized as a time slot (summation of RTS/CTS, TXOP length, and ACK). For each time slot, either one of APs wins the TXOP, or all APs that send the RTS do not receive CTS from APC, leading to Time\_Out which indicates that RTS was not approved by APC.
\end{itemize}

In Fig.\ref{fig:system}, we can have either DCF RTS/CTS or DLCA as the packet mode channel access method. If the DCF basic method is utilized, the transmission result will only be known to the AP until the TXOP duration is finished. The data load within the TXOP is much larger than the traditional scenario in which the DCF basic method is applied, leading to intolerable performance loss \cite{bianchi2000performance}.


\subsection{Maximum Achievable Data Rate}
Each AP's action $\mathbf{a}^{n}_{t}$ is directly related to the throughput of its coverage. The more TXOPs each AP gains, the higher throughput is reached. However, the number of successful TXOP contention is not the only factor to the throughput. The channel conditions between each AP and its associated STAs on different frequency bands are different, and they also vary over time. In this paper, we consider that the overall spectral efficiency of AP on its primary channel can be obtained by taking the average of the individual spectral efficiencies between AP $n$ and all the associated STAs. The channel spectral efficiency is assumed to be known by APC and denoted as $\mathbf{C}^{n}_{t} \in \mathcal{R}^{F}$ (bit/s/Hz). The spectral efficiency $\mathbf{C}^{n}_{t}(f)$ represents the maximum data that AP $n$ can achieve at time slot t on channel $f$. 



\subsection{Access Point Model}

Define a Markov decision process (MDP) for an AP over a finite state space $\mathcal{S} \in \{0,1\}^{2L} \times \mathcal{Z}$, where $L$ denotes the state size. The finite state space $\mathcal{S}$ is a set that contains concatenations of observation vectors, action vectors, and $c^{n}_{t} \in \mathcal{Z}$ that represents the total number of APs contending for TXOPs in the AP $n$'s operating frequency channel (including AP $n$ itself), i.e., 
$\mathbf{s}^{n}_{t+1}=[a^{n}_{t-L+1},o^{n}_{t-L+2},\dots,a^{n}_{t},o^{n}_{t+1},c^{n}_{t}]$. The transition function $\delta(\mathbf{s}_{t}^{n},\mathbf{s}_{t+1}^{n};a_{t}^{n})$ denotes the probability that the state $\mathbf{s}_{t}^{n}$ transfers to the state $\mathbf{s}_{t+1}^{n}$ after taking action $a_{t}^{n}$. $r(\mathbf{s}^{n}_{t},a^{n}_{t},\mathbf{s}^{n}_{t+1}) \in \mathcal{R}$ denotes the reward of AP $n$ at $t^{th}$ TXOP contention results from its state-action-state pair $\left(\mathbf{s}^{n}_{t},a^{n}_{t},\mathbf{s}^{n}_{t+1}\right)$. The accumulated discounted reward $R^{n}_{t} \in \mathcal{R}$ for AP $n$ can be expressed as
\begin{equation}
R^{n}_{t}=\sum_{k=0}^{\infty} \gamma^{k} r(\mathbf{s}^{n}_{t+k},a^{n}_{t+k},\mathbf{s}^{n}_{t+k+1}),
\end{equation}
where $\gamma \in \left(0,1\right]$ is a discounting factor. The policy of AP $n$ $\pi(n): \mathcal{S} \xrightarrow[]{} \mathcal{A}$ is assumed to be stationary, and the decision of the policy only depends on the current state. Hence, each AP aims to solve the following problem:
\begin{equation}
    \argmax_{\pi(n)} \mathbf{E}_{ \delta}[R_{t}|\mathbf{s}^{n}_{t}=\mathbf{s}, a^{n}_{t}=a, \pi(n)],
\end{equation}
which is the objective of each AP and the expectation with respect to the transition probability function $\delta$ is denoted as $E_{\delta}[\bullet]$. Since each AP simultaneously takes actions on its primary channel where each action is associated with an objective (maximization of the accumulated reward), this is overall a multi-agent problem. The calculation of the reward corresponding to various state-action pairs is detailed in the next section.

\subsection{System Reward}
Every action made by an AP has corresponding feedback (CTS/Time\_Out). The system reward is calculated as follows:
\begin{equation}\label{eq:r}
    r(\mathbf{s}^{n}_{t},a^{n}_{t},\mathbf{s}^{n}_{t+1})=\sum_{l=0}^{L-1}  \eta^{l}y^{n}_{t-l},  
\end{equation}
where $y^{n}_{t-l}=1$ denotes the successful feedback for the action $a^{n}_{t-l}$ and $y^{n}_{t-l}=-1$ for an unsuccessful contention, and $\eta \in \left[0,1\right]$ is a factor such that the more recent action is, the more weight it will have in the system reward. The overall reward estimation algorithm is shown in Appendix \ref{app:1}.

\section{DLCA Protocol}\label{sec:DRL}
In this section, we develop the following steps: a) Q-learning satisfying the Bellman optimality condition is introduced; b) we introduce deep reinforcement learning in which a deep Q network is utilized as a model for the action-value function; c) First-Order Model-Agnostic Meta-Learning (FOMAML) is applied to enhance the convergence rate and the stability of the deep Q network; d) The greedy algorithm considering PF is proposed for multi-AP coordination. 

\subsection{Q-learning}\label{AP}
In this section, we introduce standard Q-learning and $\epsilon$-greedy policy as the foundation for the following deep Q-learning. As defined in the above section, each action $a^{n}_{t}$ transfers the current state $\mathbf{s}^{n}_{t}$ of AP $n$ to $\mathbf{s}^{n}_{t+1}$ with reward $r^{n}_{t+1}$. The action-value function of AP $n$ is denoted as follows:
\begin{equation}
J(\mathbf{s}^{n}_{t},\{A^{n}_{t}\}) \triangleq  \mathbf{E}_{\delta}[\sum_{k=0}^{\infty} \gamma^{k} r(\mathbf{s}^{n}_{t+k},A^{n}_{t+k},\mathbf{s}^{n}_{t+k+1})|\mathbf{s}^{n}_{t}],
\end{equation}
where the action-value function $J(\mathbf{s}^{n}_{t},a^{n}_{t}) :\mathcal{S}\times\mathcal{A} \rightarrow{} \mathcal{R}$ outputs the accumulated reward with respect to the state $\mathbf{s}^{n}_{t}$ and the corresponding action $a^{n}_{t}$. The optimal value function is defined as:
\begin{equation}
\begin{aligned}
    V^{*}(\mathbf{s}^{n}_{t})=\max_{\mathcal{A}^{n}_{t}}J(\mathbf{s}^{n}_{t},\mathcal{A}^{n}_{t}),
\end{aligned}
\end{equation}
where $V^{*}(\mathbf{s}^{n}_{t})$ can be further written as the Bellman optimality equation:
\begin{equation}
    \begin{aligned}
    &V^{*}(\mathbf{s}^{n}_{t})=\\&\max_{a^{n}_{t} \in \mathcal{A}} \sum_{\mathbf{s}^{n}_{t+1} \in \mathcal{S}} \delta(\mathbf{s}^{n}_{t},\mathbf{s}^{n}_{t+1};{a^{n}_{t}})[r(\mathbf{s}^{n}_{t},a^{n}_{t} ,\mathbf{s}^{n}_{t+1})+\gamma V^{*}(\mathbf{s}^{n}_{t+1})],
    \end{aligned}
\end{equation}where $\delta(\mathbf{s}^{n}_{t},\mathbf{s}^{n}_{t+1};{a^{n}_{t}})$ represents the transition probability from state $\mathbf{s}^{n}_{t}$ to $\mathbf{s}^{n}_{t+1}$ after taking action $a_{t}^{n}$.
The optimal Q-function can then be expressed as the follows:
\begin{equation}
    \begin{aligned}
    & Q^{*}(\mathbf{s}^{n}_{t},a^{n}_{t}) = \\&\sum_{\mathbf{s}^{n}_{t+1} \in \mathcal{S}} \delta(\mathbf{s}^{n}_{t},\mathbf{s}^{n}_{t+1};{a^{n}_{t}})\{r(\mathbf{s}^{n}_{t},a^{n}_{t} ,\mathbf{s}^{n}_{t+1})+\gamma V^{*}(\mathbf{s}^{n}_{t+1})\},
    \end{aligned}
\end{equation}
in which the Q-function is a fixed point of a contraction operator $\mathcal{H}$ \cite{bellemare2016increasing}, defined for a generic function $Q:\mathcal{X} \times \mathcal{A} \rightarrow{} \mathcal{R}$ as the follows:
\begin{equation}
    \begin{aligned}
    (\mathcal{H}Q)(\mathbf{s}^{n}_{t},a^{n}_{t}) =&\sum_{\mathbf{s}^{n}_{t+1} \in \mathcal{S}} \delta(\mathbf{s}^{n}_{t},\mathbf{s}^{n}_{t+1};{a^{n}_{t}})\{r(\mathbf{s}^{n}_{t},a^{n}_{t} ,\mathbf{s}^{n}_{t+1}) \\ & +\gamma \max_{a^{n}_{t+1} \in \mathcal{A}} Q(\mathbf{s}^{n}_{t+1},a^{n}_{t+1})\}.
    \end{aligned}
\end{equation}
In the case of model-free reinforcement learning, the above Q-function is impossible to obtain since the transition probability is unknown. Hence, the Q-learning algorithm searches the optimal Q-function with samplings from the episodes of the MDP. Then, the Q-learning algorithm utilizes the following updating rule:
\begin{equation}\label{eq:bellman}
    \begin{aligned}
    Q(\mathbf{s}^{n}_{t},a^{n}_{t}) \leftarrow & Q(\mathbf{s}^{n}_{t},a^{n}_{t})+\beta\{r(\mathbf{s}^{n}_{t},a^{n}_{t} ,\mathbf{s}^{n}_{t+1})\\& +\gamma \max_{a^{n}_{t+1} \in \mathcal{A}}Q(\mathbf{s}^{n}_{t+1},a^{n}_{t+1})-Q(\mathbf{s}^{n}_{t},a^{n}_{t})\},
    \end{aligned}
\end{equation}
where the learning rate is denoted by $\beta$. While each AP updates $Q(\mathbf{s}^{n}_{t},a^{n}_{t})$, it also makes decisions based on $Q(\mathbf{s}^{n}_{t},a^{n}_{t})$, i.e., choosing the action corresponding to the largest Q-value. For the $\epsilon$-greedy policy, the optimal action is given by 
\begin{equation}\label{eq:greedy}
        a^{n}_{t}=
        \begin{cases}
         \text{argmax}_{a^{n}_{t}}Q(s^{n}_{t},a^{n}_{t}), &P=1-\epsilon.\\
         \text{random action}, &P= \epsilon,
        \end{cases}
\end{equation}
where $\epsilon$ denotes the probability of choosing random action. The greedy policy helps the Q-learning policy to search for more possibilities of actions randomly. It can help the policy converge faster and prevent the policy from being stuck at a sub-optimum. Q-learning is proven to converge to the optimum action-values with probability $1$ so long as all actions are repeatedly sampled in all states, and the action-values are represented discretely \cite{watkins1992q}.

\subsection{Gradient descent in Deep Q-Learning}

The traditional Q-learning algorithm can be applied to solve for the optimal policy. However, traditional Q-learning is impractical if the dimension of action-state space is large, i.e., the curse of dimensionality\cite{8103164}; thus, the well-known DQN is proposed in \cite{mnih2015human} to approximate the action-state Q-value function and the neural network used to achieve the approximation is called Q neural network (QNN).

Each AP is equipped with a QNN which outputs the approximated $Q$-value $\{Q(\mathbf{s}^{n}_{t},a^{n}_{t};\boldsymbol\theta^{n})|a^{n}_{t}\in\mathcal{A}\}$ given the input state $\mathbf{s}^{n}_{t}$ and action $a^{n}_{t}$. The optimal policy is to choose the action with the largest Q-value. Unlike the tabular update for Q-learning in Eq \eqref{eq:bellman}, the QNN in deep Q-learning can be trained by minimizing prediction errors of $Q(\mathbf{s}^{n}_{t},a^{n}_{t};\text{\boldmath$\theta$}^{n})$ at each AP and time slot, where $\text{\boldmath$\theta$}^{n}$  denotes the trainable QNN weights on AP $n$. After the reward is obtained, the state transfers to $\mathbf{s}^{n}_{t+1}$. The pair  $(\mathbf{s}^{n}_{t},a^{n}_{t},r(\mathbf{s}^{n}_{t},a^{n}_{t} ,\mathbf{s}^{n}_{t+1}),\mathbf{s}^{n}_{t+1})$ then forms a single training sample for QNN and is stored in training set $\mathcal{D}^{n}$. Please note that we will sample training data $d^{n}_{s}$ from the training set $\mathcal{D}^{n}$ for each update in the training process. Next, we define the prediction loss function of QNN as 
\begin{equation}
    L(\text{\boldmath$\theta$}^{n})=(v-Q(\mathbf{s}^{n}_{t},a^{n}_{t};\text{\boldmath$\theta$}^{n}))^2,
\end{equation}
where $Q(\mathbf{s}^{n}_{t},a^{n}_{t};\text{\boldmath$\theta$}^{n})$ is the output of QNN at time slot $t$ and the approximate value function is defined as 
\begin{equation}
    v=r(\mathbf{s}^{n}_{t},a^{n}_{t} ,\mathbf{s}^{n}_{t+1})+\gamma \max_{a^{n}_{t+1}} Q(s^{n}_{t+1},a^{n}_{t+1};\text{\boldmath$\theta$}^{n}),
    \label{eq:gam}
\end{equation}
in which the second term $\gamma \max_{a^{n}_{t+1} \in \mathcal{A}} Q(s^{n}_{t+1},a^{n}_{t+1};\text{\boldmath$\theta$})$ is obtained by searching the maximum output of QNN with respect to the selection of action $a^{n}_{t+1}$ given $s^{n}_{t+1}$. Then, we can update the trainable QNN weights using the semi-gradient algorithm \cite{sutton2018reinforcement} as below:
\begin{equation}\label{eq:semi_gradient}
\begin{aligned}
&\text{\boldmath$\theta$}^{n} \leftarrow \text{\boldmath$\theta$}^{n} +\rho\left[v-Q(s^{n}_{t},a^{n}_{t};\text{\boldmath$\theta$}^{n}) \right] \nabla Q(s^{n}_{t},a^{n}_{t};\text{\boldmath$\theta$}^{n}),
\end{aligned}
\end{equation}
where $\nabla$ is the gradient with respect to $\text{\boldmath$\theta$}^{n}$. Moreover, each AP is employed with deep Q-learning to search for the optimal QNN. However, this inevitably results in a performance loss, for some APs cannot avoid learning aggressive policies to maximize the contention benefits for themselves, and some APs learn conservative policies to avoid collision, especially when AP networks are densely overlapping. Hence, in our proposed protocol, each APs sends its QNN weights to APC that takes the average of the weights of all QNNs as follows:
\begin{equation}\label{eq:fomaml}
    \boldsymbol\theta_{g}=\frac{1}{N}\sum^{N}_{n=1} \boldsymbol\theta^{n},
\end{equation}
where $\boldsymbol\theta_{g}$ is denoted as global weight. Minimizing the above equation is equivalent to minimize the summation of all loss functions of QNNs from all APs, i.e., 
\begin{equation}
    \sum_{n=1}^{N} L(\text{\boldmath$\theta$}^{n})=\sum_{n=1}^{N}(v-Q(\mathbf{s}^{n}_{t},a^{n}_{t};\text{\boldmath$\theta$}^{n}))^2.
\end{equation}

In the meantime, minimizing the above equation implicates the following gradient descent for the summed loss function:
\begin{equation}\label{eq:global_update}
\begin{aligned}
    &\boldsymbol\theta_{g} \xleftarrow{} \boldsymbol\theta_{g} +  \frac{\rho}{N} \sum_{n=1}^{N} \left[v-Q(s^{n}_{t},a^{n}_{t};\text{\boldmath$\theta$}^{n}) \right] \nabla Q(s^{n}_{t},a^{n}_{t};\text{\boldmath$\theta$}^{n}).
\end{aligned}
\end{equation}
Hence, the global QNN weight $\boldsymbol\theta_{g}$ is equivalently obtained by iterations over the sampled data batch $d^{g}_{s}$ collected from all APs' local data set $\{\mathcal{D}^{1},\dots,\mathcal{D}^{N}\}$, which enables a faster convergence rate and lower loss function value. It is noteworthy that $\boldsymbol\theta_{g}$ is sent back to all APs from APC after global gradient descent completes according to Eq \eqref{eq:global_update}. Then,  $\boldsymbol\theta_{g}$ replaces the previous QNN weights for future training and inference. This method is called First-Order Model-Agnostic Meta-Learning (FOMAML), which can further enhance all APs' models with non-IID local data \cite{nichol2018first}.


\begin{algorithm}[t]
 \KwData{~$\boldsymbol\theta^{n}$, $\mathbf{s}_{0}^{n}$, and $t=0$.}
\While{$t \geq 0$}{

\eIf{ $mod(t,T)$ is not $T-1$ }
{
\begin{equation*}
        a^{n}_{t}=
        \begin{cases}
         \text{argmax}_{a^{n}_{t}}Q(\mathbf{s}^{n}_{t},a^{n}_{t}), &P=1-\epsilon.\\
         \text{random action}, &P= \epsilon\;
        \end{cases}
\label{eq:eps}
\end{equation*}

 Obtain $r(\mathbf{s}^{n}_{t},a^{n}_{t} ,\mathbf{s}^{n}_{t+1})$ according to Algorithm \ref{alg:mc} and store the tuple $(\mathbf{s}^{n}_{t},a^{n}_{t},r^{n}_{t},\mathbf{s}^{n}_{t+1})$ to the training batch $\mathcal{D}$\;

\If{Update}{
\vspace{3mm}
1. Sample $d^{n}_{s}$ transitions from $\mathcal{D}^{n}$.

\vspace{3mm}
2. Calculate the target value as follows:\\
$v=r^{n}_{t}+\gamma  \left(\max_{a^{n}_{t+1}}Q(\mathbf{s}^{n}_{t+1},a^{n}_{t+1};\text{\boldmath$\theta$}^{n})\right)_{Q}$.
\vspace{3mm}

3. For each randomly sample tuple in the training batch with $d_s$ samples, update $\text{\boldmath$\theta$}^{n}$ with the following gradient descent method:
\begin{equation*}
    \boldsymbol\theta^{n} \xleftarrow{} \boldsymbol\theta^{n}  -\rho \nabla_{\text{\boldmath$\theta$}^{n}}L(\text{\boldmath$\theta$}^{n}).
\end{equation*}}


}{
4. Each AP send its QNN weights to APC that obtains the global QNN $\boldsymbol\theta_{g}=\frac{1}{N}\sum^{N}_{n=1} \boldsymbol\theta_{n}$. 

\vspace{3mm}

5. Sample $d_{s}^{g}$ from $\{\mathcal{D}^{1}, \dots, \mathcal{D}^{N}\}$ and update $\boldsymbol\theta_{g}$ with the following gradient descent method: 
\begin{equation*}
\begin{aligned}
    &\boldsymbol\theta_{g} \xleftarrow{} \boldsymbol\theta_{g} + \\& \frac{\rho}{N} \sum_{n=1}^{N} \left[v-Q(s^{n}_{t},a^{n}_{t};\text{\boldmath$\theta$}^{n}) \right] \nabla Q(s^{n}_{t},a^{n}_{t};\text{\boldmath$\theta$}^{n}).
\end{aligned}
\end{equation*}

\vspace{3mm}

6. Send $\boldsymbol\theta_{g}$ back to each AP: $\boldsymbol\theta^{n} \leftarrow \boldsymbol\theta_{g}$.
}
$t=t+1$\;
}
\caption{DLCA Algorithm.}
\label{alg:ql}
\end{algorithm}

\begin{remark}
In Algorithm \ref{alg:ql}, samples are collected from the local data set in step 1. Then, in step 2, target value is calculated. Gradient descent method is implemented in step 3 based on the collected samples and the calculated target value. Step 4, 5, and 6 represent the FOMAML method and is triggered once every $T$ local training loops. It is noteworthy that the Q-value in step 2 can be obtained by looping the action corresponding to the largest Q-value with time complexity of $\mathcal{O}(F)$. The gradient in machine learning is normally computed using the back-propagation method \cite{goodfellow2016deep} as a numerical solution with time complexity of $\mathcal{O}(FM)$. Algorithm \ref{alg:ql} is typically executed in batch mode - such that QNN update occurs once per batch to reduce computation load. FOMAML is only triggered every period of $T$ to reduce the communication overhead between the APC and AP. The global QNN $\boldsymbol\theta_{g}$ is trained on APC using global information gathered by APC, i.e., $\{d_{s}^{1}, \dots, d_{s}^{N}\}$.
\end{remark}


\subsection{AP Coordination: Greedy Algorithm}\label{PF}
In the above section, each AP runs with a deep Q-learning algorithm independently. However, the channel on which each AP should run is not described. In this section, the APC policy that allocates channels to all APs considering PF is proposed.

Denote $\phi^{n}_{t}(f)$ as the instantaneous proportional achievable data rate for AP $n$ at time slot $t$ at channel $f$. We assume the block fading channel condition to explore the convergence property of our proposed algorithm, i.e., $\mathbf{C}^{n}_{t}(f) = \mathbf{C}^{n}(f)$ is assumed to be constant over multiple TXOP slots. Then we have $\phi^{n}(f)=\frac{\mathbf{C}^{n}(f)}{\mathbf{n}(f)}$. Denote $\mathbf{x}_{t}^{n}(f)$ as the actual data rate of AP $n$ at time slot $t$ at channel $f$, the allocation scheme for AP $n$ can then be expressed as follows:
\begin{equation}\label{eq:PF}
    f^{*} = \argmax_{f} P_{t}^{n}(f),
\end{equation}
where
\begin{equation}
P_{t}^{n}(f) = \frac{\phi^{n}(f)}{  \tilde{D}_{t}^{n} }
\end{equation}
in which
\begin{equation}
\tilde{D}_{t}^{n}= \left( 1 - \frac{1}{t} \right)\tilde{D}_{t-1}^{n}+\frac{\mathbf{1}^{T}\mathbf{x}_{t}^{n}}{t},
\end{equation}
and $\tilde{D}_{t}^{n}$ represents the average throughput of AP $n$ up to time slot $t$. Note that only one element in $\mathbf{x}_{t}^{n} \in \mathcal{R}^{F}$ is non-zero. After allocating one channel to an AP, the scheduler updates the ratio $P_{t}^{n}(f)$ for the next AP's allocation. The proposed greedy algorithm considering PF \cite{srikant2013communication} on APC is specified in Algorithm 2. In Algorithm 2, each AP is allocated with a channel in the while-loop. The average throughput of AP $n$ up to time slot $t$ is calculated. Then, the channel is chosen to maintain the current PF. The greedy algorithm can guarantee the asymptotic PF, and the corresponding proof is shown in Appendix \ref{pf_proof}.
\begin{algorithm}[t]
\caption{Greedy Algorithm Considering PF on APC.} 
\label{greedy}
\KwData{$x^{n}_{0}$, $\tilde{D}^{n}_{0}$, $C^{n}(f)$, and $n=0$.}

$\mathbf{n}(f)=0$ for all $f$\;

 \While{$n \leq N$}{

\begin{equation*}\label{eq:greedy_step1}
\tilde{D}_{t}^{n} \xleftarrow{} \left( 1 - \frac{1}{t} \right)\tilde{D}_{t-1}^{n}+\frac{\mathbf{1}^{T}\mathbf{x}_{t}^{n}}{t}.
\end{equation*}

\begin{equation*}\label{eq:greedy_step2}
  f^{*} \xleftarrow{} \arg{\max\left\{ \frac{\phi^{n}(f)}{\tilde{D}^{n}_{t}} \right\}}
\end{equation*}
where 
\begin{equation*}
\phi^{n}(f)=\frac{\mathbf{C}^{n}(f)}{\mathbf{n}(f)}.
\end{equation*}
$n \xleftarrow{} n+1$\; $\mathbf{n}(f) \xleftarrow{} \mathbf{n}(f)+1$\; Allocate AP n to $f^{th}$ channel\;

}

\label{alg:greedy}
\end{algorithm}

\begin{figure}[t]
    \centering
    \includegraphics[width = .42\textwidth,height=.25\textwidth]{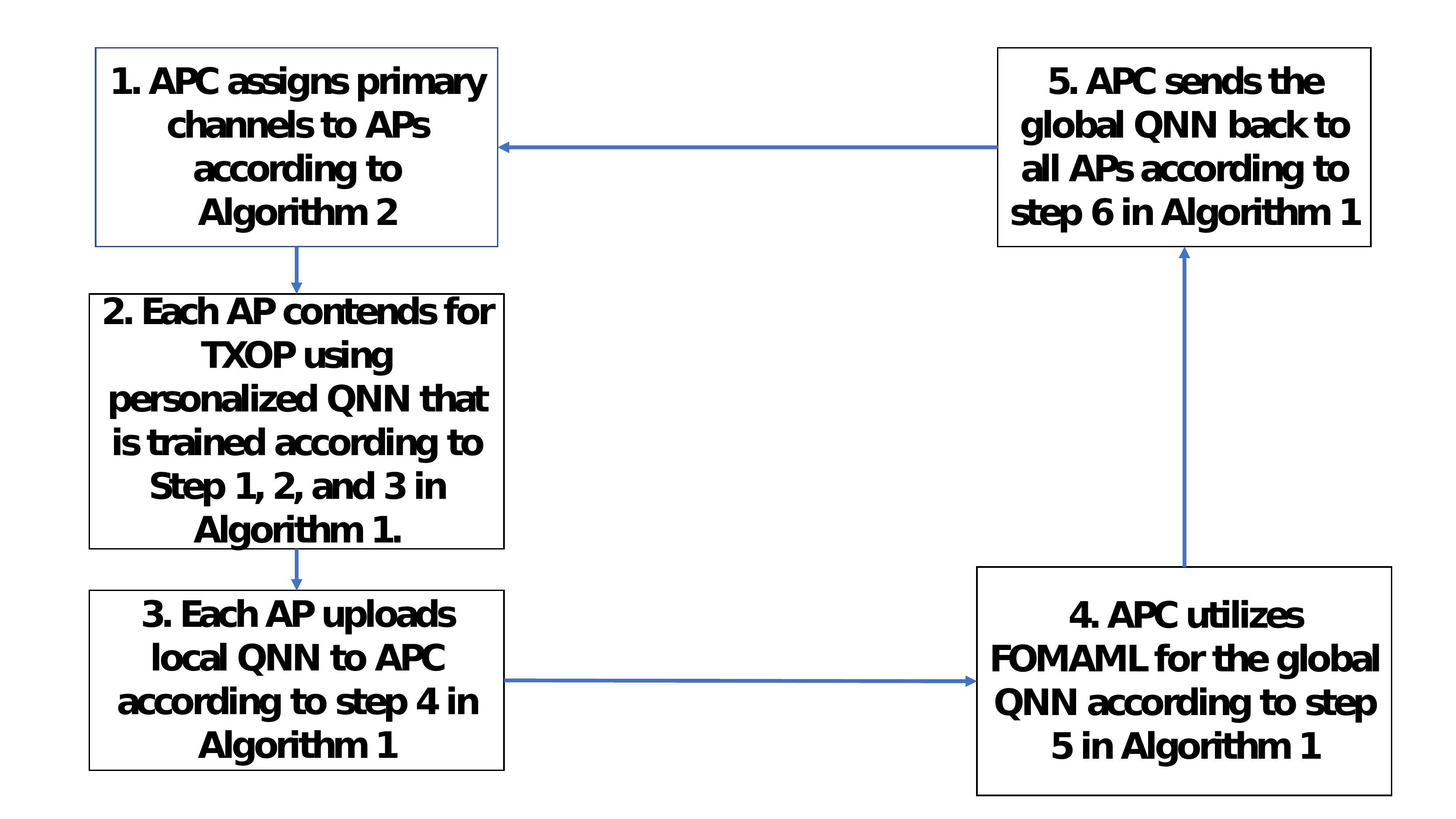}
    \caption{Flow Chart of DLCA Protocol: DLCA + Greedy Algorithm + FOMAML.}
    \label{fig:flowchart}
\end{figure}

Round Robin (RR) and PF have been developed as two common scheduling strategies. Among those, PF is widely considered in wireless networks. Different AP has different average throughput due to the number of previously gained TXOPs and various channel spectral efficiency. The greedy algorithm considering PF exploits these variations by allocating the primary channel to the AP with the best conditions for the upcoming TXOP slot. As a design approach, this approach is superior to RR. In the end, the proposed DLCA protocol is shown in Fig.\ref{fig:flowchart}.

\begin{remark}
In Algorithm \ref{alg:greedy}, the computation complexity of the PF scheduling method is $\mathcal{O}(NF)$. Similar to FOMAML, The greedy algorithm is only triggered every period of $T$ on APC to reduce the communication burden between the APC and AP. The information of instantaneous data rate $\mathbf{x}^{n}_{t}$, average data rate $\tilde{D}_{t}^{n}$, and spectral efficiency $\mathbf{C}^{n}(f)$ can be exchanged between APs and APC through the wired connection such as Light Weight Access Point Protocol (LWAPP), wireless connection such as DCF. FOMAML and the greedy algorithm triggered with a suitable period has negligible impact on the network throughput as long as the QNN is of lightweight.
\end{remark}

\section{Performance Evaluation}
\label{simulation}
In this section, we present simulation results for a dense overlapping network that implements our DLCA protocol. The performance comparison between the DLCA protocol, SH-TXOP, and DCF RTS/CTS for the overlapping network characterizes the superiority of our DLCA protocol. In the end, we evaluate proportional fairness and stability achieved by the greedy algorithm in the DLCA protocol.

\subsection{Throughput in Multi-AP and Multi-band Networks}
With the FCC opening up the 6 GHz \cite{Fcc} band for unlicensed use for 5G wireless networks, joint operation in $5$ and $6$ GHz is feasible with orthogonal sub-channels with a bandwidth of $20$ MHz. Our simulations consist of a fully overlapping multi-AP network using $5$ GHz and $6$ GHz bands- for a total of $F$ sub-channels of $20$ MHz. The TXOP slot is granted to the shared APs as a multiple of $32$~$\mu s$, and the maximum amount of time granted is $8.16$~ms. The TXOP is thus set as $8.16$ ms. In the simulation, AP is assumed to be operating in the saturation mode, i.e., it is always necessary for AP to gain TXOPs because AP needs to communicate with its associated STAs in common Wi-Fi networks continuously. The average value of the spectral efficiency on each channel is $40$ Mbps \cite{Perez2019AP}, and the spectral efficiency is assumed to be an uniform distribution, i.e., $\mathbf{C}^{n}_{t}(f) \sim \mathbf{U}[1, 3]~\text{bit/s/Hz}$. FOMAML mechanism and the greedy algorithm are triggered with the period of $T=100$ ms. We conduct Monte-Carlo simulations with 100 independent trials and then take the average of the results.


We firstly consider SH-TXOP without APC \cite{ahn2020novel} as the baseline. In the SH-TXOP protocol, APs share the universal frequency band rather than operating on different sub-channels. After a certain AP wins the wide-band TXOP, it starts to share the TXOP on each 20 MHz sub-channel to other APs based on round robin (RR) method, indicated by Announcement Trigger frame (ATF) that contains the information such as the channel allocation scheme for shared APs. For example, if there are 16 APs and 4 channels and AP $1$ wins the TXOP, then AP $1$ allocates the first sub-channel as the primary channel for itself. Next, it allocates the second sub-channel to AP $2$, and etc. Each AP contends for the wide-band TXOP using IEEE 802.11 DCF basic \cite{bianchi2000performance}. The Backoff Window size of DCF basic is $CW_{min}=32$ and $m=6$. Note that DCF basic method can be utilized in this scenario. Since the packet size in the process of gaining the sharing opportunity of TXOP is negligible, DCF basic is applicable in such a scenario. However, DCF basic is well known to perform worse than DCF RTS/CTS \cite{bianchi2000performance}. To futher enhance the DCF RTS/CTS, a model is proposed in \cite{Gao2017TCOM} that increases the network throughput by optimizing the initial backoff window size for DCF RTS/CTS.  For this simulation, all APs are equally allocated to a fixed primary channel at the beginning. For example, if there are 16 APs and 4 channels, then AP $1-4$ are allocated to the first channel as their primary channel, and AP $5-8$ are allocated to the second channel as their primary channel and so on. Next, each AP contends for TXOP in its primary channel using DCF RTS/CTS with the optimized initial backoff window size \cite{Gao2017TCOM}. This method in the simulation is called RTS/CTS. The aggregate network throughput $x$ of the entire network with $N$ APs for this method is expressed as follows:
\begin{equation}
    x = \sum^N_{n=1}x^{n}=\sum^N_{n=1}\frac{z^{n}LU}{L+\delta},
    \label{eq:agrtpt}
\end{equation}
where $L$ is the information bits in one packet, $\delta=\delta_{0}U$ stands for the protocol overhead in the unit of bits. The channel bit rate $U$ can be further written as the product of the sub-channel bandwidth and the link spectral efficiency. The above equation is further explained in Appendix \ref{sec:RTS}. The system parameters are summarized in Table \ref{tb:sim_pra}. 



\begin{table}[]
\caption{System Parameters for Multi-AP Networks}
\label{tb:sim_pra}
\centering
\resizebox{.27\textwidth}{!}{%
\begin{tabular}{|c|c|}
\hline
\textbf{Parameters} &\textbf{Value} \\                                  \hline
slot time ($\mu$s)                        & $50$    \\     \hline
SIFS ($\mu$s)                        & 28                                   \\ \hline
DIFS ($\mu$s)                        & 128                                   \\ \hline
PHY Header ($\mu$s)                  & 20                                   \\ \hline
TXOP ($\mu$s)                  &  640                                \\ \hline
CTS\_Timeout ($\mu$s)                 & 300 
 \\ \hline 
ACK\_Timeout ($\mu$s)                 & 300 
 \\ \hline
Headers (Bytes)                      & 36                                   \\ \hline
ACK (Bytes)                          & 14 + PHY Header                      \\ \hline
RTS (Bytes)                          & 20 + PHY Header                      \\ \hline
CTS (Bytes)                          & 14 + PHY Header                      \\ \hline
ATF (Bytes)                          & 16 + PHY Header
                  \\ \hline
\end{tabular}%
} 
\end{table}

\begin{figure}[h]
    \centering
    \includegraphics[height=.35\textwidth]{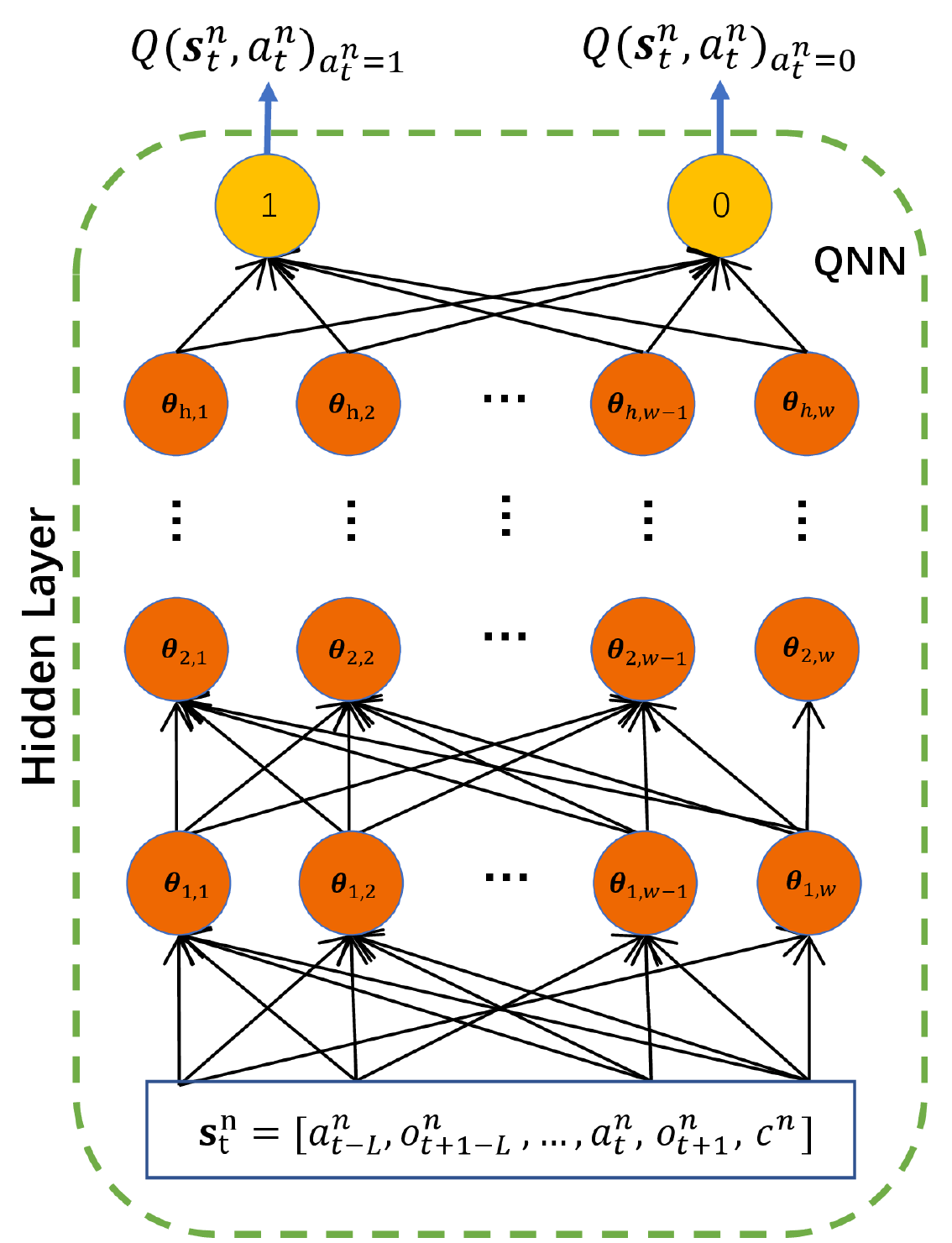}
    \caption{QNN: Fully-connected Neural Network.}
    \label{fig:dqn}
\end{figure}

We utilize PyTorch \cite{NEURIPS2019_9015} to train QNN for DLCA. The simulations are conducted on a server with a CPU (Intel Core i7-9700k) and a GPU (NVIDIA GeForce GTX 2080Ti) in Python language. QNN is constructed by $h=5$ fully connected layers with $64$ neurons in each layer, which is illustrated in Fig.\ref{fig:dqn}. The operation of the QNN starts by taking the state vector as the input. Then, it outputs two Q values corresponding to action - transmission and action - wait respectively. AP decides to transmit if the corresponding Q value is larger and wait otherwise. Table \ref{tb:hyper_pra} lists the hyper-parameter of the deep Q-learning. ReLU (Rectified Linear Unit) defined as
\begin{equation}
    f(x)=x^{+}=\max(0,x)
\end{equation}
is utilized as the activation function to the input of each neuron in the QNN. Unlike other activation functions such as the sigmoid function, ReLu can help the QNN avoid the vanishing gradient issue because the gradient of $f(x)$ when $x > 0$ is always a constant. Therefore, choosing ReLu can prompt faster learning process and better performance. 
\begin{table}[h]
\caption{Hyper-Parameters of QNN}
\label{tb:hyper_pra}
\centering
\resizebox{.28\textwidth}{!}{
\begin{tabular}{|c|c|}
\hline
\textbf{Parameters}         & \textbf{Value}  \\ \hline
State size & 40 \\ \hline
Batch size & 32 \\ \hline
Learning rate $\rho$ & 0.001   \\ \hline
$\gamma$ in Eq \eqref{eq:gam} & 0.9 \\ \hline
$\epsilon$ in Eq \eqref{eq:greedy} &  0.05 \\ \hline
$\eta$ in Eq \eqref{eq:r} &  0.5\\ \hline
Step size of $\lambda$ & 0.1 \\ \hline
Gradient descent step size $u$ & 0.25 \\ \hline

\end{tabular}
}
\end{table}

\begin{figure*}[h] 
  \centering
  
  \subfigure[The Number of Channel is F=4] {\includegraphics[width=.3\textwidth,height=0.15\textwidth]{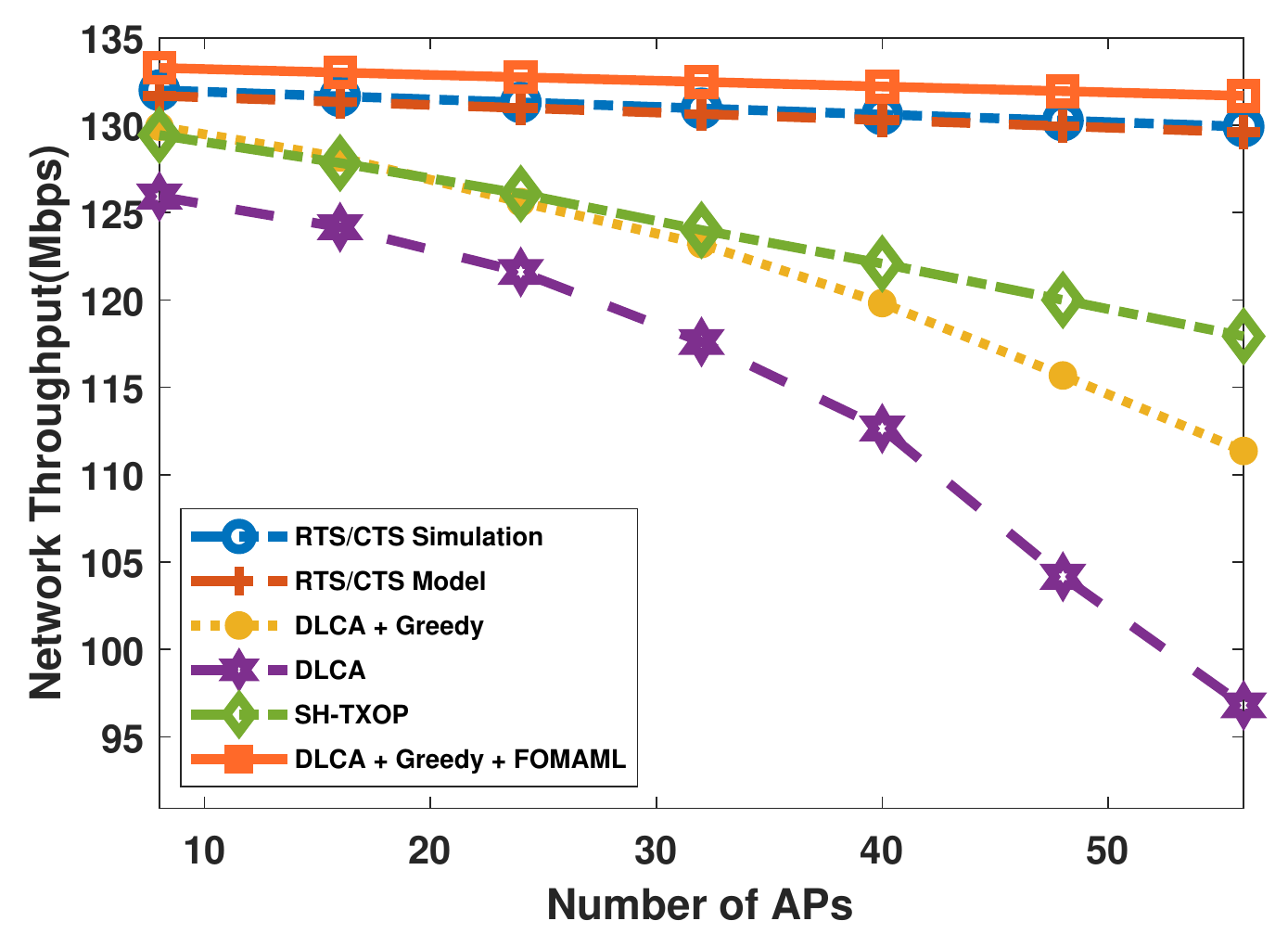}\label{fig:performance2}}
  \quad
  \subfigure[The Number of Channel is F=8] {\includegraphics[width=.3\textwidth,height=0.15\textwidth]{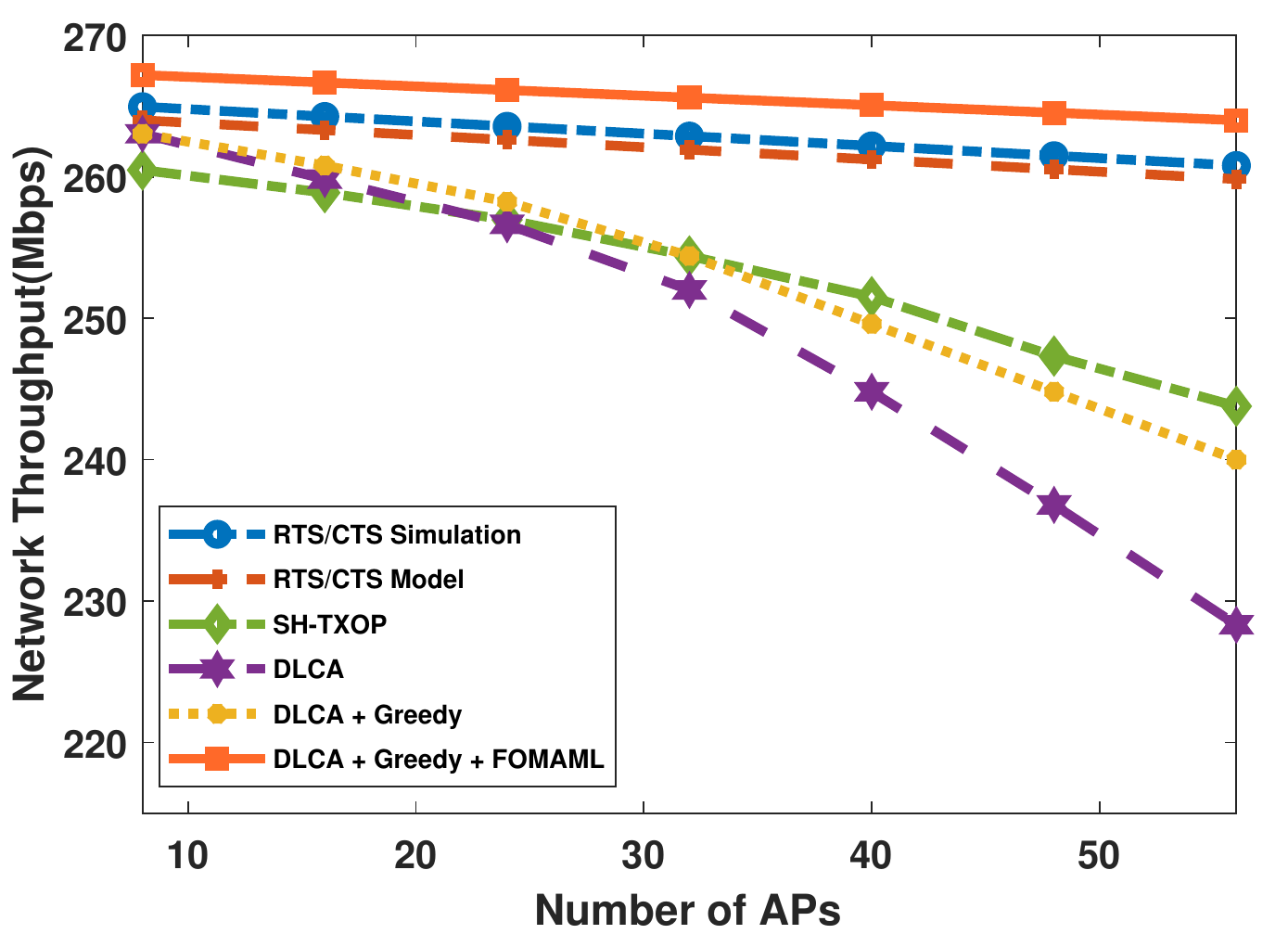}\label{fig:performance1}}
  \quad
    \subfigure[The Number of Channel is F=16] {\includegraphics[width=.3\textwidth,height=0.15\textwidth]{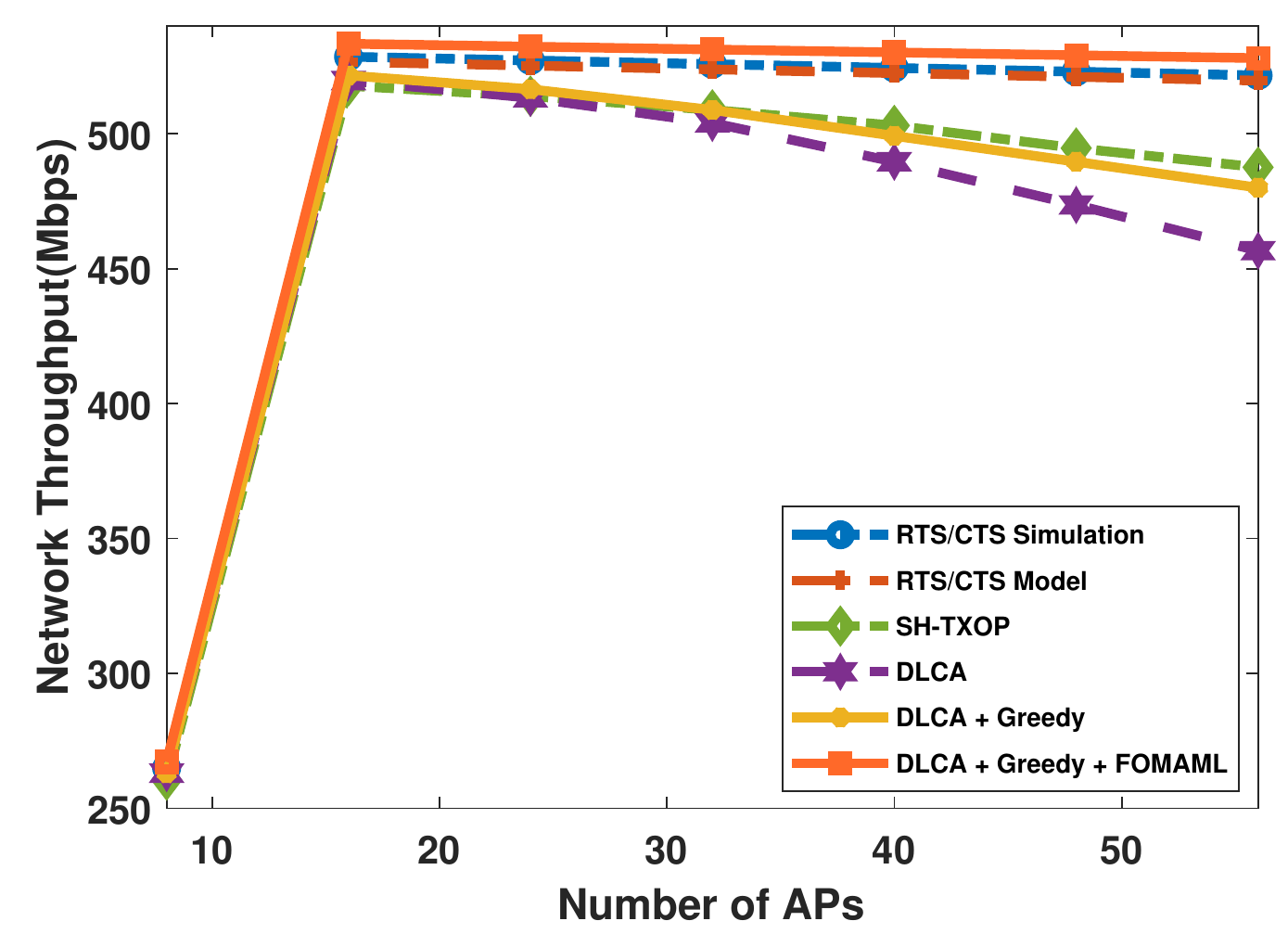}\label{fig:performance3}}\quad
  
  \caption{Network Throughput vs Number of APs. }
  
  \label{fig:performance}
\end{figure*}

\begin{figure}[h]
    \centering
    \includegraphics[width=.35\textwidth,height=.2\textwidth]{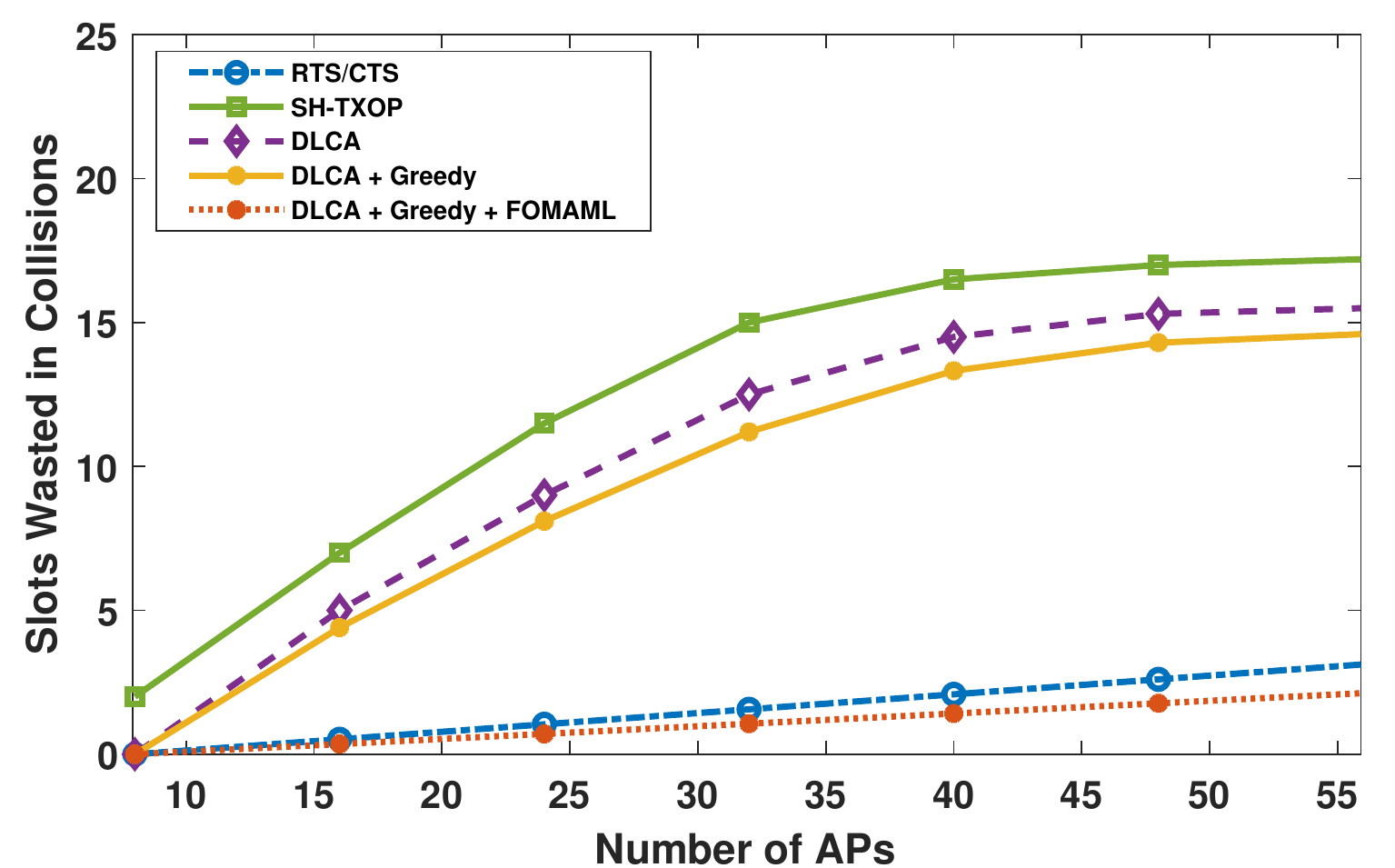}
    \caption{Average Collision vs Number of APs. The Number of Channel is $F=8$.}
    \label{fig:collision}
\end{figure}

\begin{figure}[h]
    \centering
    \includegraphics[width=.35\textwidth,height=.2\textwidth]{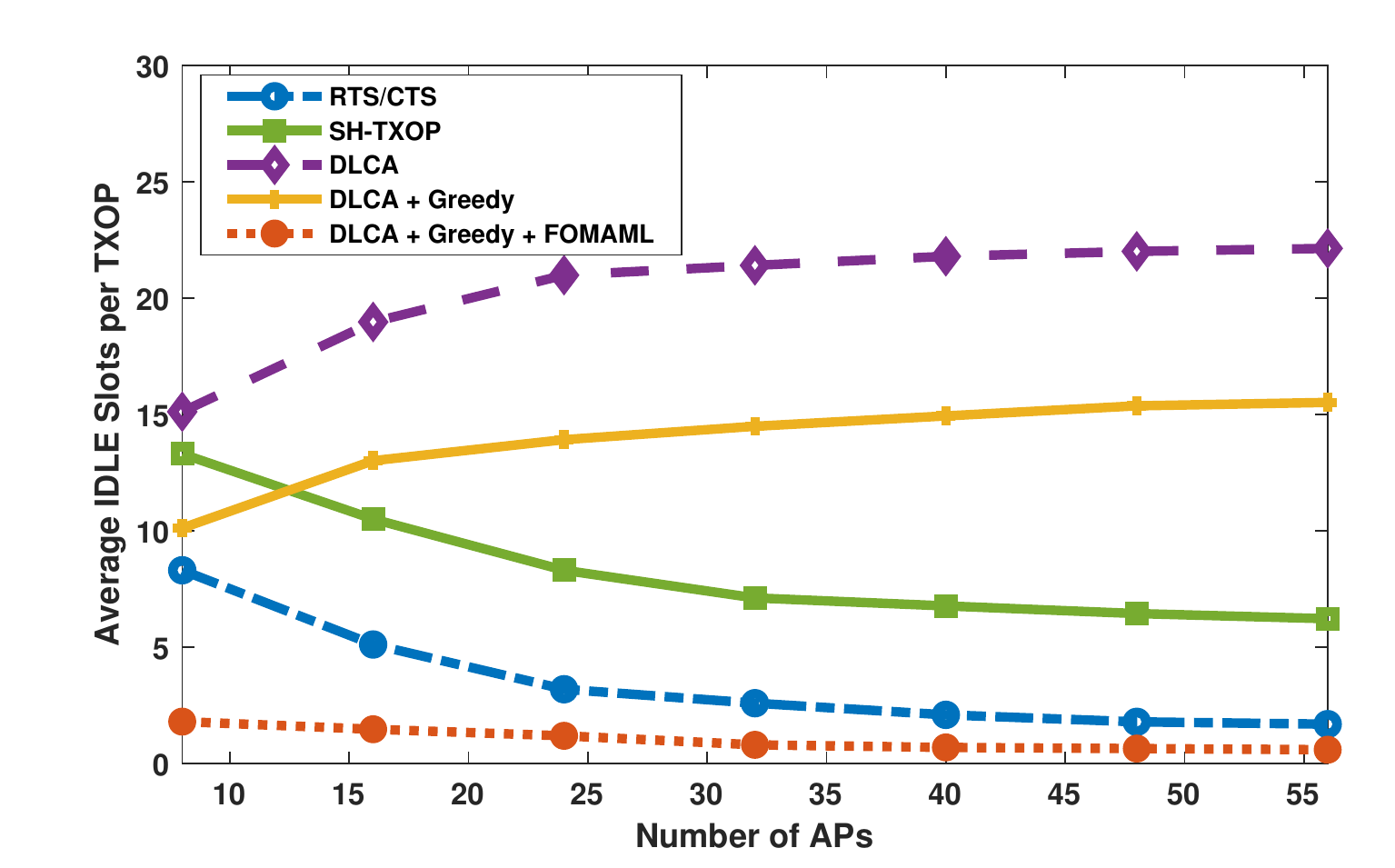}
    \caption{Average IDLE vs Number of APs. The Number of Channel is $F=8$.}
    \label{fig:idle}
\end{figure}

\begin{figure}[h]
    \centering
    \includegraphics[width=.35\textwidth, height=.16\textwidth]{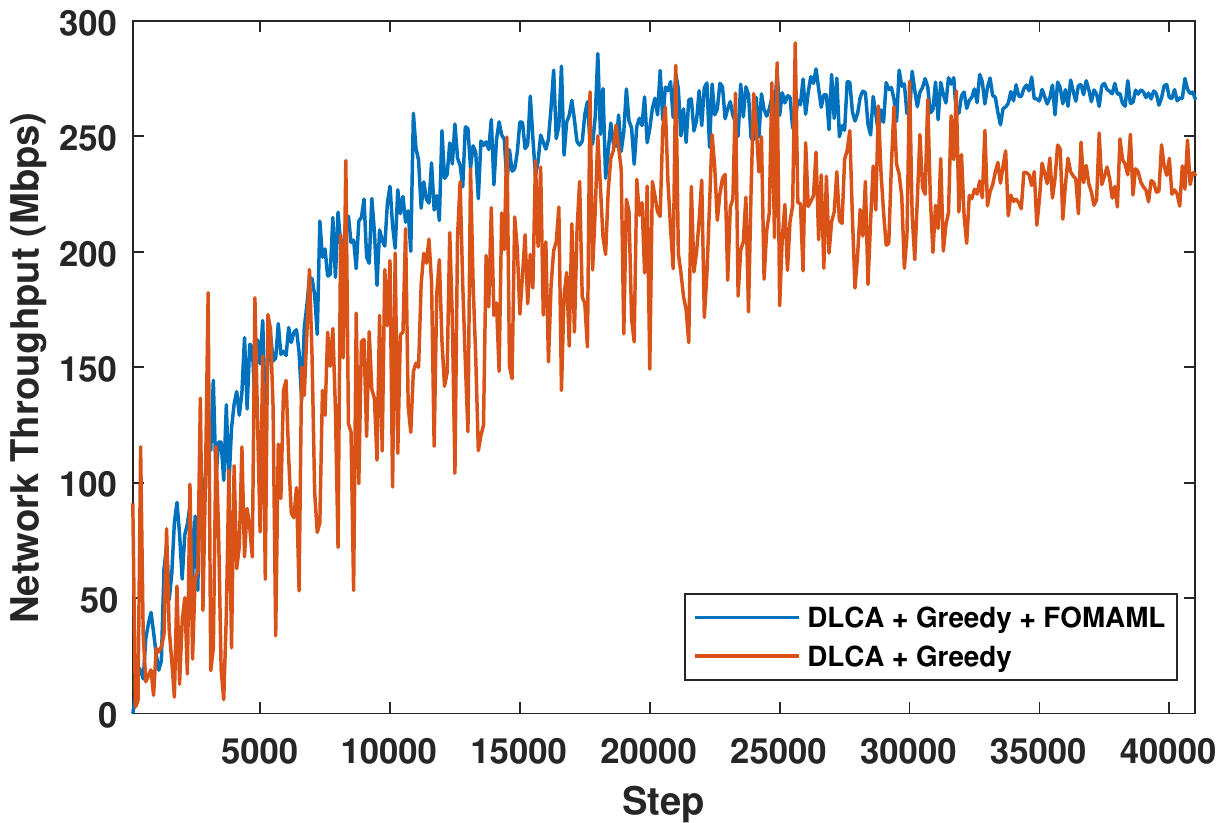}
    \caption{Convergence: Throughput vs Training Steps. The Number of Channel $F=8$, The Number of APs $N=16$. Each time step is one round of gradient descent in Eq. \eqref{eq:semi_gradient}. }
    \label{fig:loss}
\end{figure}

The simulation result of the network throughput is shown in Fig.\ref{fig:performance}. We validate the precision of the RTS/CTS model by showing that the simulation results are close to their corresponding theoretical results. Three curves related to the DLCA protocol are plotted respectively. The curve labeled with DLCA reflects the simulation of running distributed deep reinforcement learning on each AP without primary channel allocation from the APC. In this case, each AP only has a fixed primary channel allocation similar to the RTS/CTS simulation. On the other hand, DLCA + greedy method represents the simulation of running distributed deep reinforcement learning on each AP with primary channel allocation decision from APC. Note that APC does not update QNN globally in DLCA + greedy method. In Fig.\ref{fig:performance1}, when the number of APs is small, the performance of DLCA and DLCA + greedy method is better than SH-TXOP. The reason is that when the number of APs is $N=8$ for both DLCA methods, each AP has its own exclusive primary channel and no collision happens at all, which can be observed from Fig.\ref{fig:collision}. However, 8 APs have to contend for the wide-band TXOP for SH-TXOP. The average slots wasted in the collision for DLCA and DLCA + greedy method are zero, not to mention that SH-TXOP has more average IDLE slots per TXOP than DLCA + greedy. As the number of APs increases, SH-TXOP outperforms DLCA and DLCA + greedy. This is because some APs can develop very aggressive TXOP contention policy without the supervision from FOMAML. Note that each AP only wants to maximize its own total reward. Hence, it is possible that APs assigned to one channel are all aggressive and we can view this training process as Prisoner's Dilemma; that is, if one AP does not develop aggressive TXOP contention policy, then it has no chance to get any TXOP forever. Therefore, it is likely that DLCA will lead to high collision probability. However, SH-TXOP and DCF RTS/CTS have backoff counter to avoid collision probability if collision happens. Hence, as the number of APs increases, average slots wasted in collisions and average IDLE slots per TXOP all increases for DLCA and DLCA + greedy in Fig.\ref{fig:collision} and \ref{fig:idle}, which leads to severe performance loss. Although both RTS/CTS method and DLCA + greedy + FOMAML perform well, DLCA + greedy + FOMAML provides higher network throughput than RTS/CTS method by 3\% for the total number of AP ranging from $8$ to $56$. This is because the overhead remains in RTS/CTS and the backoff window takes up time slots without sending any data packet. One can observe the RTS/CTS method has more average collision slots and IDLE slots from Fig.\ref{fig:collision} and \ref{fig:idle}. In Fig.\ref{fig:performance3}, for the case of 16 channels 
with 8 APs, only half of the channel resources are utilized. Hence, the network throughput grows linear with increasing number of APs at the beginning. In Fig.\ref{fig:performance}, DLCA + Greedy + FOMAML outperforms SH-TXOP by 10\% when the number of APs is $N=56$ in average of three cases. The advantage of FOMAML can also be demonstrated in Fig.\ref{fig:loss}, during the training process, the network throughput in both methods have large variance in the initial learning phase. However, DLCA + greedy + FOMAML has faster convergence rate and smaller variance. This can be attributed to the fact that each AP's self-training only reaches local optimality, emphasizing the important role of FOMAML as the global optimizer that achieves the necessary AP coordination for throughput maximization.  




  

\subsection{PF in Multi-AP and Multi-band Networks}

In the above section, aggregate network throughput is simulated. However, aggregate network throughput does not reflect the throughput of each AP, leading to a potential issue that the maximum throughput can always be achieved by having the same AP holding the channel, and no fairness exists at all. Hence, fairness must be guaranteed so that each AP in the network can utilize the TXOP for UL/DL communication with associated STAs. The greedy algorithm has been proven to enable PF among APs asymptotically previously. This section implements simulations to study a network utility metric that describes PF and network throughput. Meanwhile, the stability of our proposed algorithm is also investigated.

\begin{figure}[h]
    \centering
    \includegraphics[width=.36\textwidth,height=.18\textwidth]{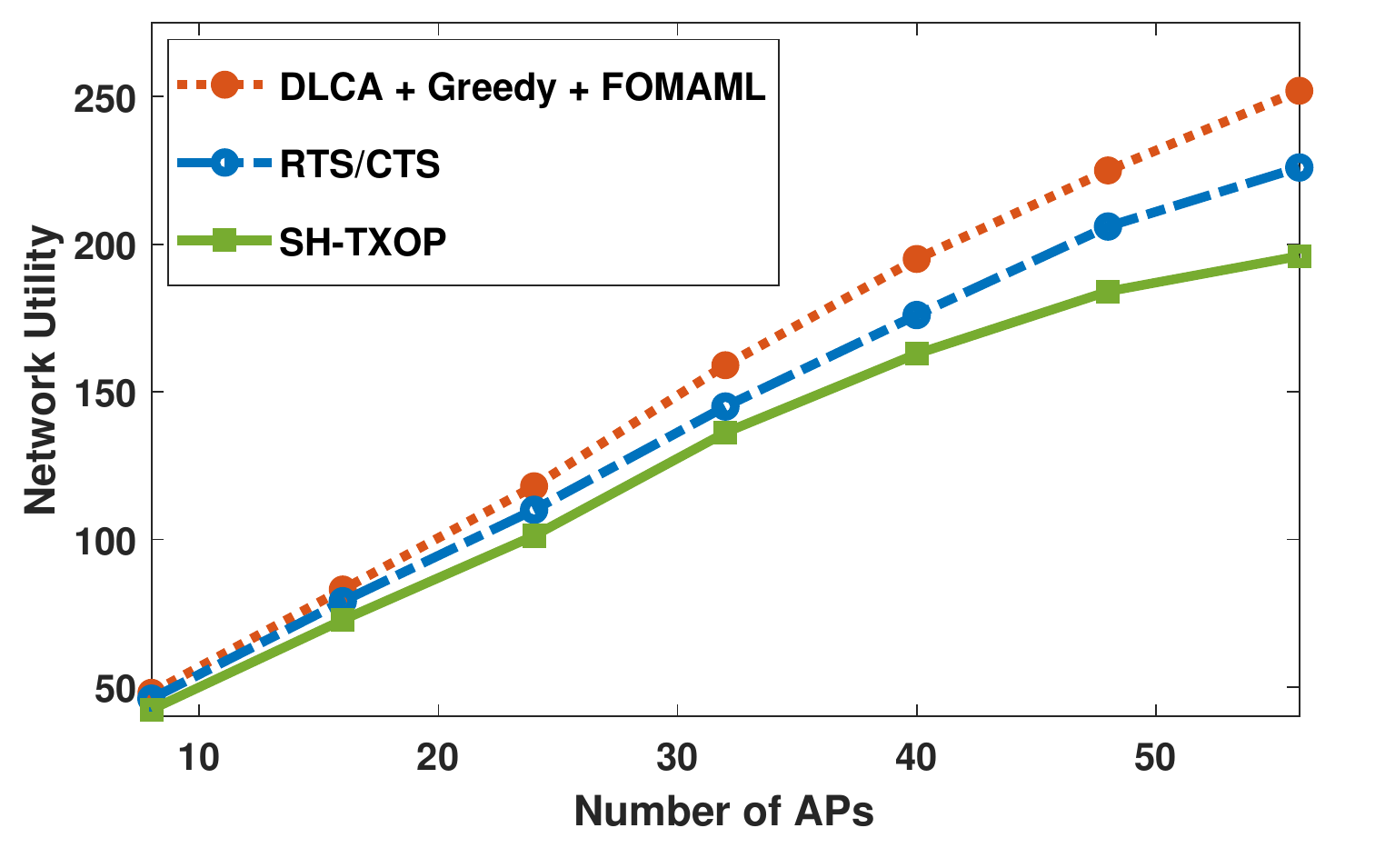}
    \caption{Network Utility ($\sum_{n=1}^{N}\log(\bar{D}^{n})$) vs Number of APs. The number of channel is $F=8$.}
    \label{fig:utility}
\end{figure}

\begin{figure}[h]
    \centering
    \includegraphics[width=.36\textwidth, height=.18\textwidth]{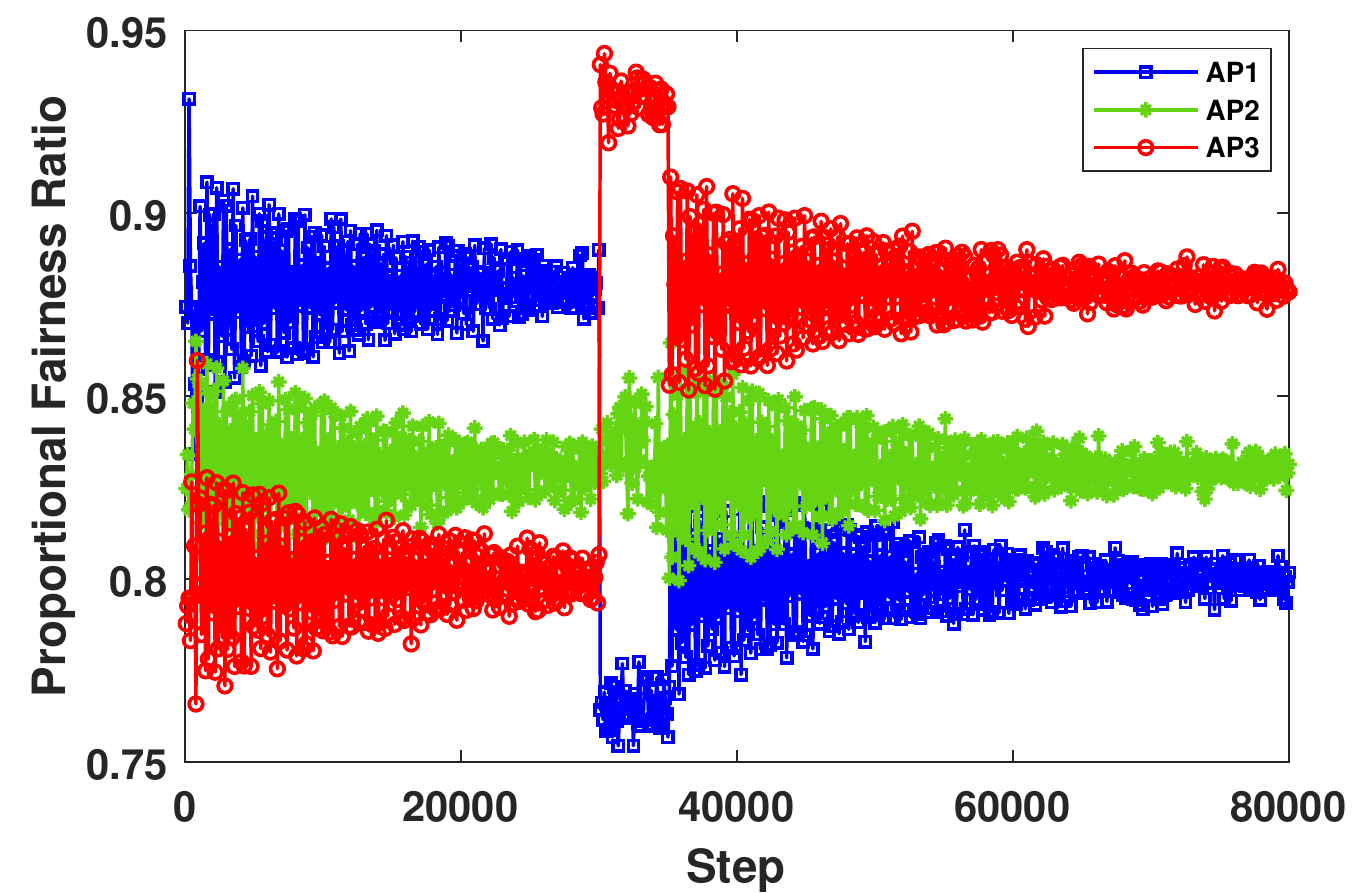}
    \caption{PF ratio $b^{n}$ vs Time Step with Greedy Algorithm. PF ratios of $3$ APs out of $N=18$ APs on $F=8$ channels are depicted. Each time step is one slot time in Tab. \ref{tb:sim_pra}.}
    \label{fig:pf}
\end{figure}

To be consistent with the notations in Appendix.\ref{pf_proof}, the network utility is defined as $\sum_{n=1}^{N}\log(\bar{D}^{n})$,
where $\bar{D}^{n}$ is the average data rate of AP $n$. We simulate the network utility of DLCA + greedy + FOMAML, SH-TXOP. and RTS/CTS for comparison. In Fig.\ref{fig:utility}, DLCA protocol performs better than RTS/CTS by 13.8\% and SH-TXOP by 28.3\% when the number of APs equals $N=56$. In 
$\sum_{n=1}^{N}\log(\bar{D}^{n})$, the $log$ term punishes the AP that has low throughput. Therefore, the network utility demonstrates joint network throughput and PF. We next show the stability of DLCA + greedy + FOMAML in terms of AP's PF ratio, which is expressed as
\begin{equation}
    b^{n}=\frac{\bar{D^n}}{\phi^n}.
\end{equation}
According to Eq \eqref{eq:PF_alg}, the closer $b^{n}$ of each AP is to each other, the better PF is achieved. In Fig.\ref{fig:pf}, the PF ratios of 3 APs converge to $0.87$, $0.84$, and $0.81$ respectively after $30000$ steps. Then, we exchange the value of spectral efficiency $\mathbf{C}^{n}_{t}(f)$ between AP 1 and AP 3 at step $30000$ to demonstrate the stability. After a sudden change at step 30000, three curves experience drastic oscillation. Then, we can observe that the curves of AP 1 and AP 3 converge again eventually. This result indicates that the fairness broken by a sudden network change can be restored very quickly. Hence, our greedy algorithm is shown to be robust and efficient.

In the end, we conclude that the design of multi-AP network with APC has a higher upper limit than the multi-AP network without APC in terms of the network throughput. The choice of RTS/CTS or DLCA + greedy + FOMAML is constrained by the hardware and energy cost. For the AP with limited power constraint, one can choose RTS/CTS with lower power consumption but the performance loss, especially the lack of PF, might lead to an unsatisfied user experience. On the other hand, if the power budget is high enough and each AP is able to run light-weight QNN with suitable CPU or GPU, AP can reach higher network throughput, and proportional fairness among APs can be also guaranteed.

\section{Conclusion}
\label{sec:conclusion}

In this paper, we propose enhancements to the RRM architecture for dense overlapping Wi-Fi networks that align with the proposed coordinated AP operation in Wi-Fi 7 (802.11be). Specifically, we develop a novel multi-AP coordination system architecture with DLCA protocol. The proposed protocol considers not only the network throughput maximization but also the proportional fairness among APs. The performance of DLCA related algorithms is then evaluated via simulations and compared with SH-TXOP protocol and RTS/CTS as benchmarks. The numerical results show that DLCA outperforms SH-TXOP and state-of-the-art RTS/CTS with an optimized initial back-off window in terms of network throughput and network utility. Moreover, convergence rate and stability are also demonstrated in the simulation. 

This paper studies the fully overlapping dense Wi-Fi networks. In our future work, we will investigate partially overlapping dense Wi-Fi networks. In such a case, optimization for dynamic AP coordination set, frequency reuse in different coordination set for 802.11be, and coexistence with other protocols will be considered.  




\appendix

\subsection{Reward Estimation Method}\label{app:1}
Algorithm \ref{alg:mc} is a modified version of Monte Carlo method in \cite{sutton2018reinforcement} that aims to estimate reward and help Q-learning converges faster. For AP $n$ operating in $f^{th}$ frequency channel, we initialize the total reward to be zero in step 1. Then, if the current action $a_{t}^{n}(f)=1$, we use while loop to find all feedback of the transmission action (action value is equal to $1$) in the state vector. Suppose one of the feedback is ACK, we add one to the weighted total reward value in step 2 since it is a successful transmission. Otherwise, we minus one to punish the weighted total reward value in step 3. Suppose the current action $a_{t}^{n}(f)=0$, then the total reward value is $1$ if busy channel is sensed in step 4 since the AP successfully avoids a potential collision. The total reward value is set to $-1$ as a punishment if idle channel is sensed in step 5.

\begin{algorithm}[h]
 \KwData{$f$, $\mathbf{s}_{t}^{n}$, feedback for all actions in $\mathbf{s}_{t}^{n}$.}
 \KwResult{$r(\mathbf{s}^{n}_{t},a^{n}_{t},\mathbf{s}^{n}_{t+1})$}

 1. $r(\mathbf{s}^{n}_{t},a^{n}_{t},\mathbf{s}^{n}_{t+1})\leftarrow{}\mathbf{0}$\;

 \eIf{$a^{n}_{t}(f)==\{1\}$}{
 $l \leftarrow{} L$\;
 \While{$l \geq 0$}{
 \eIf{the feedback of $a^{n}_{t-l}(f) = 1$ is ACK}{
\begin{equation*}\label{eq:reward_plus}
    2.~r(\mathbf{s}^{n}_{t},a^{n}_{t},\mathbf{s}^{n}_{t+1})\leftarrow{}\eta \times r(\mathbf{s}^{n}_{t},a^{n}_{t},\mathbf{s}^{n}_{t+1})+
1
\end{equation*} 
 } {
 \begin{equation*}\label{eq:reward_minus}
     3.~r(\mathbf{s}^{n}_{t},a^{n}_{t},\mathbf{s}^{n}_{t+1})\leftarrow{}\eta \times r(\mathbf{s}^{n}_{t},a^{n}_{t},\mathbf{s}^{n}_{t+1})-
1
 \end{equation*}
 }

  $l\leftarrow{}l-1$\;
 }

}
 {

 \eIf{$\mathbf{o}^{n}_{t+1}(f)==\{1\}$}{
 \begin{equation*}\label{eq:observe_plus}
     4.~r(\mathbf{s}^{n}_{t},a^{n}_{t},\mathbf{s}^{n}_{t+1})\leftarrow{}1
 \end{equation*}}
 {
 \begin{equation*}\label{eq:observe_minus}
    ~~~ 5.~r(\mathbf{s}^{n}_{t},a^{n}_{t},\mathbf{s}^{n}_{t+1})\leftarrow{}-1
 \end{equation*}
 }
 
 }
 \caption{Reward estimation method.}
 \label{alg:mc}
\end{algorithm}

\subsection{Proof of Asymptotic Proportional Fairness}\label{pf_proof}
We show that the greedy algorithm considering PF maximizes the aggregate throughput while guaranteeing the asymptotic PF on one channel. Denote $p_{t}^{n}(f)$ as the probability of the AP $n$ at time slot $t$ being assigned with channel $f$. Assume the spectral efficiency does not change within $t$ slots, then we have 
\begin{equation}
    \bar{D}^{n}_{t} =  \sum_{f} \sum_t p_{t}^{n}(f)\phi^{n}(f).
\end{equation}

Therefore, the network utility maximization problem \cite{srikant2013communication} is
\begin{align}
&\max ~ \sum_{n}{\log( \sum_{f} \sum_t p_{t}^{n}(f)\phi^{n}(f))} \label{eq:utility} \\
&\text{s.t.} ~\sum_{n}\sum_{f} p_{t}^{n}(f) \leq 1,
p_{t}^{n}(f) \geq 0, ~\forall n, f, t.
\end{align}
Note that the above problem is a convex problem since log function with composition of an affine function still preserves concavity. Applying Lagrange multipliers, we obtain the following:
\begin{equation}
    \begin{aligned}
    \sum_{n}{\log( \sum_{t} \sum_{f} p_{t}^{n}(f)\phi^{n}(f))}- \sum_{t} \lambda_{t}(\sum_{n}\sum_{f} p_{t}^{n}(f)-1)
    \end{aligned}
\end{equation}

Taking the derivative w.r.t. $p_{t}^{n}(f)$, we obtain the optimal solution as 

\begin{equation}
    \frac{\phi^{n}}{\bar{D}^{n}_{t}} - \lambda^{*} = 0 \quad \text{if} \quad p_t^{n}(f) > 0,
\end{equation}
Asymptotically, the PF algorithm helps the AP network to reach the PF, i.e., 
\begin{equation}\label{eq:PF_alg}
    \lim_{t \xrightarrow{} \infty}\frac{\bar{D}_{t}^{1} }{\phi^{1}} = \dots = \lim_{t \xrightarrow{} \infty}\frac{\bar{D}_{t}^{n} }{\phi^{n}}.
\end{equation}
Hence, allocation method shown in Eq. \eqref{eq:PF} follows the optimal condition of the 11be network utility maximization problem with PF.

\subsection{Performance of Multi-AP IEEE 802.11 RTS/CTS Networks}
\label{sec:RTS}
\begin{figure}[h]
    \centering
    \includegraphics[width = .5\textwidth]{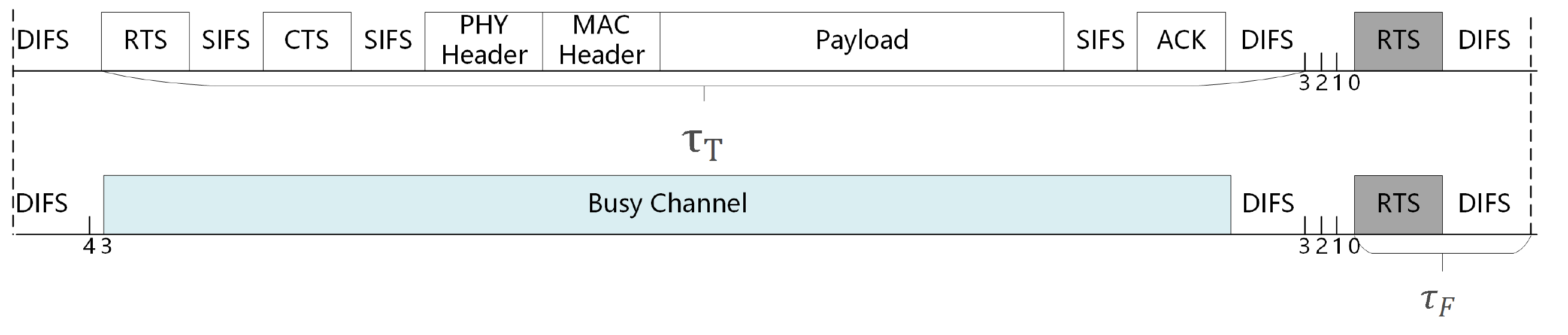}
    \caption{Graphic illustration of successful transmission and collision in DCF networks with the RTS/CTS access mechanism.}
    \label{fig:csma_protocol}
\end{figure}
\begin{figure}[h]
    \centering
    \includegraphics[width = .5\textwidth]{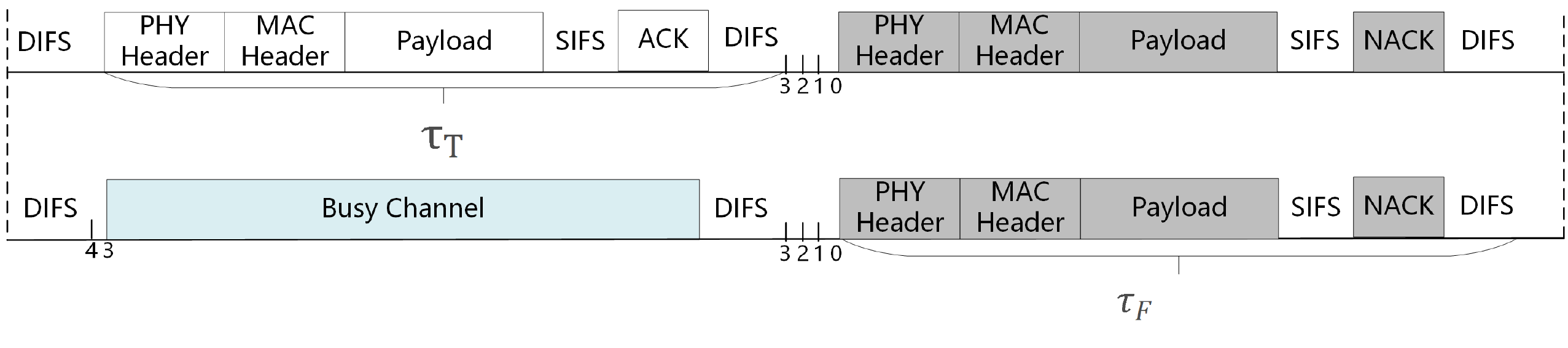}
    \caption{Graphic illustration of successful transmission and collision with DCF basic mechanism.}
    \label{fig:dcf_basic}
\end{figure}
The aggregate network data rate $x$ is the average number of information bits successfully transmitted per second, which is the sum of the average number of information bits that AP $n$ successfully transmits per second $x^n$
\begin{equation}
    x=\sum_{n=1}^{N}x^{n},
\end{equation}
The duration to transmit a packet consisting of $L$ information bits is given by $\frac{L}{U}+\delta_{0}$, where $U$ is the channel bit rate, and $\delta_{0}$ is the protocol overhead in seconds. With the RTS/CTS mechanism and DCF basic illustrated in Fig. \ref{fig:csma_protocol} and \ref{fig:dcf_basic} respectively, $\delta_{0}$ is given by 
\begin{equation}
\begin{aligned}
        \delta_{0}^{RTS}&=\frac{\text{RTS}+\text{CTS}+\text{ACK}}{U_{b}}+\text{Header}+\text{DIFS}+3\times \text{SIFS}\\
    \delta_{0}^{basic}&=\frac{\text{ACK}}{U_{b}}+\text{Header}+\text{DIFS}+ \text{SIFS},
\end{aligned}
\end{equation}
where $U_{b}$ denotes the basic rate. In the end, the data rate $x^{n}$ of each BSS $i$ can be expressed as follows:
\begin{equation*}
    x^{n}=\frac{L z^{n}}{\frac{L}{U}+\delta_{0}}=\frac{z^{n}LU}{L+\delta},
\end{equation*}
where $\delta=\delta_{0}U$ stands for the protocol overhead in the unit of bits. The channel bit rate $U$ can be further written as the product of the channel bandwidth of each AP and the link spectral efficiency. The details of derivation of $z^{n}$ is related to parameters of $\tau_{T}$ and $\tau_{F}$ \cite{Gao2017TCOM}. As for the throughput of Multi-AP IEEE 802.11 DCF basic network without APC, $\tau_{T}^{DCF} = \tau_{T} - (RTS+CTS+2\times SIFS)$. $\tau_{F}^{DCF}$ is equal to $\tau_{T}^{DCF}$ in basic DCF since there is no RTS/CTS and the collision leads to a waste of whole packet time instead of RTS/CTS with short duration, which is the reason why RTS/CTS can enhance the system throughput.

\bibliographystyle{IEEEtran}
\bibliography{ref}

\begin{thebibliography}{10}
\providecommand{\url}[1]{#1}
\csname url@samestyle\endcsname
\providecommand{\newblock}{\relax}
\providecommand{\bibinfo}[2]{#2}
\providecommand{\BIBentrySTDinterwordspacing}{\spaceskip=0pt\relax}
\providecommand{\BIBentryALTinterwordstretchfactor}{4}
\providecommand{\BIBentryALTinterwordspacing}{\spaceskip=\fontdimen2\font plus
\BIBentryALTinterwordstretchfactor\fontdimen3\font minus
  \fontdimen4\font\relax}
\providecommand{\BIBforeignlanguage}[2]{{%
\expandafter\ifx\csname l@#1\endcsname\relax
\typeout{** WARNING: IEEEtran.bst: No hyphenation pattern has been}%
\typeout{** loaded for the language `#1'. Using the pattern for}%
\typeout{** the default language instead.}%
\else
\language=\csname l@#1\endcsname
\fi
#2}}
\providecommand{\BIBdecl}{\relax}
\BIBdecl

\bibitem{ZhangVTC2020}
L.~{Zhang}, H.~{Yin}, Z.~{Zhou}, S.~{Roy}, and Y.~{Sun}, ``Enhancing {WiFi}
  multiple access performance with federated deep reinforcement learning,'' in
  \emph{IEEE 92nd Vehicular Technology Conference (VTC2020-Fall)}.

\bibitem{Cisco}
Cisco, ``{Cisco Annual Internet Report (2018–2023)},'' \emph{White Paper},
  Cisco System Inc., Mar. 2020.

\bibitem{Perez2019AP}
D.~Lopez-Perez, A.~Garcia-Rodriguez, L.~Galati-Giordano, M.~Kasslin, and
  K.~Doppler, ``{IEEE 802.11be Extremely High Throughput: The Next Generation
  of Wi-Fi Technology Beyond 802.11ax},'' \emph{IEEE Communications Magazine},
  vol.~57, no.~9, pp. 113--119, Sept. 2019.

\bibitem{11axTutorial}
E.~{Khorov}, A.~{Kiryanov}, A.~{Lyakhov}, and G.~{Bianchi}, ``{A Tutorial on
  IEEE 802.11ax High Efficiency WLANs},'' \emph{IEEE Communications Surveys
  Tutorials}, vol.~21, no.~1, pp. 197--216, Sep. 2019.

\bibitem{11beMultiAP}
J.~{Liu}, T.~{Pare}, Y.~{Seok}, J.~{Wang}, F.~{Hsu}, and J.~{Yee},
  ``{Multi-{AP} Enhancement and {Multi-Band} Operations},'' Mediatek Inc.,
  Tech. Rep. IEEE 802.11-18/1155r1, Jun. 2018.

\bibitem{shtxop}
S.~Naribole, W.~B. Lee, K.~Srinivas, R.~Duan, and A.~Ranganath, ``{Shared TXOP
  Protocol},'' Samsung Inc., Tech. Rep. IEEE 802.11-20/0277r1, Mar. 2020.

\bibitem{kwak2005performance}
B.~{Kwak}, N.~{Song}, and L.~{Miller}, ``{Performance Analysis of Exponential
  Backoff},'' \emph{IEEE/ACM Transactions on Networking}, vol.~13, no.~2, pp.
  343--355, Apr. 2005.

\bibitem{bianchi2000performance}
G.~Bianchi, ``{Performance Analysis of the {IEEE} 802.11 Distributed
  Coordination Function},'' \emph{{IEEE Journal on Selected Areas in
  Communications}}, vol.~18, no.~3, pp. 535--547, Mar. 2000.

\bibitem{ahn2020novel}
W.~Ahn, ``{Novel Multi-AP Coordinated Transmission Scheme for 7th Generation
  WLAN 802.11 be},'' \emph{Entropy}, vol.~22, no.~12, p. 1426, 2020.

\bibitem{7727996}
L.~Sequeira, J.~L. de~la Cruz, J.~Ruiz-Mas, J.~Saldana, J.~Fernandez-Navajas,
  and J.~Almodovar, ``{Building an SDN Enterprise WLAN Based on Virtual APs},''
  \emph{IEEE Communications Letters}, vol.~21, no.~2, pp. 374--377, 2017.

\bibitem{application1}
Y.~Wang, X.~Li, P.~Wan, and R.~Shao, ``{Intelligent Dynamic Spectrum Access
  Using Deep Reinforcement Learning for VANETs},'' \emph{IEEE Sensors Journal},
  vol.~21, no.~14, pp. 15\,554--15\,563, Feb. 2021.

\bibitem{application2}
H.~Song, L.~Liu, J.~Ashdown, and Y.~Yi, ``{A Deep Reinforcement Learning
  Framework for Spectrum Management in Dynamic Spectrum Access},'' \emph{IEEE
  Internet of Things Journal}, vol.~8, no.~14, pp. 11\,208--11\,218, Jan. 2021.

\bibitem{application3}
W.~Ahsan, W.~Yi, Z.~Qin, Y.~Liu, and A.~Nallanathan, ``{Resource Allocation in
  Uplink NOMA-IoT Networks: A Reinforcement-Learning Approach},'' \emph{IEEE
  Transactions on Wireless Communications}, vol.~20, no.~8, pp. 5083--5098,
  Mar. 2021.

\bibitem{ding2020deep}
H.~Ding, F.~Zhao, J.~Tian, D.~Li, and H.~Zhang, ``{A Deep Reinforcement
  Learning for User Association and Power Control in Heterogeneous Networks},''
  \emph{Ad Hoc Networks}, vol. 102, p. 102069, May 2020.

\bibitem{mnih2015human}
V.~Mnih, K.~Kavukcuoglu, D.~Silver, A.~A. Rusu, J.~Veness, M.~G. Bellemare,
  A.~Graves, M.~Riedmiller, A.~K. Fidjeland, G.~Ostrovski \emph{et~al.},
  ``{Human-level Control through Deep Reinforcement Learning},'' \emph{Nature},
  vol. 518, no. 7540, pp. 529--533, Feb. 2015.

\bibitem{kihira2020adversarial}
Y.~Kihira, Y.~Koda, K.~Yamamoto, T.~Nishio, and M.~Morikura, ``{Adversarial
  Reinforcement Learning-based Robust Access Point Coordination against
  Uncoordinated Interference},'' \emph{arXiv preprint arXiv:2004.00835}, 2020.

\bibitem{yu2019deep}
Y.~{Yu}, T.~{Wang}, and S.~{Liew}, ``{Deep-reinforcement Learning Multiple
  Access for Heterogeneous Wireless Networks},'' \emph{IEEE Journal on Selected
  Areas in Communications}, vol.~37, no.~6, pp. 1277--1290, Mar. 2019.

\bibitem{yu2020non}
Y.~Yu, S.~C. Liew, and T.~Wang, ``{Non-uniform Time-step Deep {Q-network} for
  Carrier-sense Multiple Access in Heterogeneous Wireless Networks},''
  \emph{IEEE Transactions on Mobile Computing}, Apr. 2020.

\bibitem{yu20211}
------, ``{Multi-Agent Deep Reinforcement Learning Multiple Access for
  Heterogeneous Wireless Networks with Imperfect Channels},'' \emph{IEEE
  Transactions on Mobile Computing}, pp. 1--1, Feb. 2021.

\bibitem{yu20212}
X.~Ye, Y.~Yu, and L.~Fu, ``{Multi-channel Opportunistic Access for
  Heterogeneous Networks Based on Deep Reinforcement Learning},'' \emph{IEEE
  Transactions on Wireless Communications}, pp. 1--1, July 2021.

\bibitem{bellemare2016increasing}
M.~G. Bellemare, G.~Ostrovski, A.~Guez, P.~Thomas, and R.~Munos, ``Increasing
  the action gap: New operators for reinforcement learning,'' in
  \emph{Proceedings of the AAAI Conference on Artificial Intelligence},
  vol.~30, no.~1, 2016.

\bibitem{watkins1992q}
C.~{Watkins} and P.~{Dayan}, ``{Technical Note: Q-Learning},'' \emph{Machine
  learning}, vol.~8, no. 3-4, pp. 279--292, May 1992.

\bibitem{8103164}
K.~Arulkumaran, M.~P. Deisenroth, M.~Brundage, and A.~A. Bharath, ``{Deep
  Reinforcement Learning: A Brief Survey},'' \emph{IEEE Signal Processing
  Magazine}, vol.~34, no.~6, pp. 26--38, Nov. 2017.

\bibitem{sutton2018reinforcement}
R.~{Sutton} and A.~{Barto}, \emph{{Reinforcement learning: An
  introduction}}.\hskip 1em plus 0.5em minus 0.4em\relax MIT press, 2018.

\bibitem{nichol2018first}
A.~Fallah, A.~Mokhtari, and A.~Ozdaglar, ``On the convergence theory of
  gradient-based model-agnostic meta-learning algorithms,'' in
  \emph{International Conference on Artificial Intelligence and
  Statistics}.\hskip 1em plus 0.5em minus 0.4em\relax PMLR, 2020, pp.
  1082--1092.

\bibitem{goodfellow2016deep}
I.~{Goodfellow}, Y.~{Bengio}, and A.~{Courville}, \emph{{Deep Learning}}, MIT
  press, 2016.

\bibitem{srikant2013communication}
S.~{Rayadurgam} and Y.~{Lei}, \emph{{Communication Networks: An Optimization,
  Control, and Stochastic Networks Perspective}}, Cambridge University Press,
  2013.

\bibitem{Fcc}
{Federal Communications Commission (FCC)}, ``{Unlicensed Use of the 6 {GHz}
  Band},'' \emph{Docket No. 17-183}, Apr. 2020.

\bibitem{Gao2017TCOM}
Y.~{Gao}, L.~{Dai}, and X.~{Hei}, ``{Throughput Optimization of Multi-BSS IEEE
  802.11 Networks With Universal Frequency Reuse},'' \emph{IEEE Transactions on
  Communications}, vol.~65, no.~8, pp. 3399--3414, May 2017.

\bibitem{NEURIPS2019_9015}
A.~Paszke, S.~Gross, F.~Massa, A.~Lerer, J.~Bradbury, G.~Chanan, T.~Killeen,
  Z.~Lin, N.~Gimelshein, L.~Antiga \emph{et~al.}, ``{Pytorch: An Imperative
  Style, High-performance Deep Learning Library},'' in \emph{Advances in Neural
  Information Processing Systems}, 2019, pp. 8026--8037.

\end{thebibliography}
\end{document}